\documentstyle[twoside,12pt,epsf]{article}
\input epsf
\epsfverbosetrue
\def\singlespace {\smallskipamount=3pt plus1pt minus1pt
                  \medskipamount=6pt plus2pt minus2pt
                  \bigskipamount=12pt plus4pt minus4pt
                  \normalbaselineskip=12pt plus0pt minus0pt
                  \normallineskip=1pt
                  \normallineskiplimit=0pt
                  \jot=3pt
                  {\def\smallskip {\vskip\smallskipamount}}
                  {\def\medskip   {\vskip\medskipamount}}
                  {\def\bigskip   {\vskip\bigskipamount}}
                  {\setbox\strutbox=\hbox{\vrule
                    height8.5pt depth3.5pt width 0pt}}
                  \parskip 0pt
                  \normalbaselines}
\def\doublespace {\smallskipamount=6pt plus2pt minus2pt
                  \medskipamount=12pt plus4pt minus4pt
                  \bigskipamount=24pt plus8pt minus8pt
                  \normalbaselineskip=24pt plus0pt minus0pt
                  \normallineskip=2pt
                  \normallineskiplimit=0pt
                  \jot=6pt
                  {\def\smallskip {\vskip\smallskipamount}}
                  {\def\medskip   {\vskip\medskipamount}}
                  {\def\bigskip   {\vskip\bigskipamount}}
                  {\setbox\strutbox=\hbox{\vrule
                    height17.0pt depth7.0pt width 0pt}}
                  \parskip 12.0pt
                  \normalbaselines}
\def\halfspace {\smallskipamount=6pt plus2pt minus2pt
                  \medskipamount=12pt plus4pt minus4pt
                  \bigskipamount=24pt plus8pt minus8pt
                  \normalbaselineskip=16pt plus0pt minus0pt
                  \normallineskip=2pt
                  \normallineskiplimit=0pt
                  \jot=6pt
                  {\def\smallskip {\vskip\smallskipamount}}
                  {\def\medskip   {\vskip\medskipamount}}
                  {\def\bigskip   {\vskip\bigskipamount}}
                  {\setbox\strutbox=\hbox{\vrule
                    height17.0pt depth7.0pt width 0pt}}
                  \parskip 12.0pt
                  \normalbaselines}

\def\pprintspace {\smallskipamount=4pt plus1pt minus1pt
                  \medskipamount=9pt plus2pt minus2pt
                  \bigskipamount=16pt plus4pt minus4pt
                  \normalbaselineskip=14pt plus0pt minus0pt
                  \normallineskip=1pt
                  \normallineskiplimit=0pt
                  \jot=4pt
                  {\def\smallskip {\vskip\smallskipamount}}
                  {\def\medskip   {\vskip\medskipamount}}
                  {\def\bigskip   {\vskip\bigskipamount}}
                  {\setbox\strutbox=\hbox{\vrule
                   height9.5pt depth4.5pt width 0pt}}
                  \parskip 0pt
                  \normalbaselines}
\def\reidelspace {\smallskipamount=.1667 true in plus4pt minus2pt
                  \medskipamount=.3333 true in plus8pt minus2pt
                  \bigskipamount=13 true pt plus2pt minus2pt
                  \normalbaselineskip=13 true pt plus0pt minus0pt
                  \normallineskip=1 true pt
                  \normallineskiplimit=0 true pt
                  \jot=3pt
                  {\def\smallskip {\vskip\smallskipamount}}
                  {\def\medskip   {\vskip\medskipamount}}
                  {\def\bigskip   {\vskip\bigskipamount}}
                  {\setbox\strutbox=\hbox{\vrule
                    height8.5pt depth3.5pt width 0pt}}
                  \parskip 0pt
                  \normalbaselines}

\def\folio{\ifnum\pageno=1\nopagenumbers\else\number\pageno\fi}

\def\refitem{\par\noindent\hangindent 20pt}

\def\wisk#1{\ifmmode{#1}\else{$#1$}\fi}

\def\lt     {\wisk{<}}
\def\gt     {\wisk{>}}

\def\lsim   {\wisk{_<\atop^{\sim}}}

\def\amin   {\wisk{^{\prime}}}

\def\deg    {\wisk{^\circ}}
\def\ddeg   {\wisk{{\rlap.}^\circ}}

\def\thin{\thinspace}
\def\muK{\wisk{{\rm \mu K}}}

\begin{document}
\pagestyle{plain}
\pprintspace

\large
\begin{center}
Calibration and Systematic Error Analysis \\
For the {\it COBE\thin}\footnotemark[1] DMR Four-Year Sky Maps
\end{center}
\noindent \footnotetext[1]
{
~The National Aeronautics and Space Administration/Goddard Space Flight Center 
(NASA/GSFC) is responsible for the design, development, and operation of the
Cosmic Background Explorer ({\it COBE}).  Scientific guidance is provided by 
the {\it COBE} Science Working Group.  
}

\medskip
\normalsize
\pprintspace
\noindent
\begin{center}
A.~Kogut\footnotemark[2]$^{,3}$,
A.J. Banday$^{2,4}$,
C.L. Bennett$^5$,
K.M. G\'{o}rski$^{2,6}$,
G. Hinshaw$^2$, \\
P.D. Jackson$^2$,
P. Keegstra$^2$,
C. Lineweaver$^7$,
G.F. Smoot$^7$,
L. Tenorio$^7$,
and
E.L. Wright$^8$
\end{center}
\footnotetext[2]{
~Hughes STX Corporation, Laboratory for Astronomy and Solar Physics, 
Code 685, NASA/GSFC, Greenbelt MD 20771. \newline
\indent~$^3$ E-mail: kogut@stars.gsfc.nasa.gov. \newline
\indent~$^4$ Current address: Max Planck Institut f\"{u}r Astrophysik,
85740 Garching Bei M\"{u}nchen, Germany. \newline
\indent~$^5$ Laboratory for Astronomy and Solar Physics, 
NASA Goddard Space Flight Center, Code 685, Greenbelt MD 20771. \newline
\indent~$^6$ On leave from Warsaw University Observatory,
Aleje Ujazdowskie 4, 00-478 Warszawa, Poland. \newline
\indent~$^7$ Lawrence Berkeley Laboratory, Building 50-25,
University of California at Berkeley,
Berkeley, CA 90024. \newline
\indent~$^8$ UCLA Astronomy, PO Box 951562, Los Angeles, CA 90095-1562. \newline
}

\medskip
\normalsize
\pprintspace
\begin{center}
{\it COBE} Preprint 96-10 \\
Submitted to {\it The Astrophysical Journal} \\
January 5, 1996\\
\end{center}


\medskip
\begin{center}
\large
ABSTRACT
\end{center}

\normalsize
\noindent
The Differential Microwave Radiometers (DMR) instrument
aboard the {\it Cosmic Background Explorer (COBE)} 
has mapped the full microwave sky 
to mean sensitivity 26 \muK ~per 7\deg ~field of view.
The absolute calibration is determined to 0.7\% 
with drifts smaller than 0.2\% per year.
We have analyzed both the raw differential data and the pixelized sky maps
for evidence of contaminating sources
such as solar system foregrounds,
instrumental susceptibilities,
and artifacts from data recovery and processing.
Most systematic effects couple only weakly to the sky maps.
The largest uncertainties in the maps result from 
the instrument susceptibility to the Earth's magnetic field,
microwave emission from the Earth,
and upper limits to potential effects at the spacecraft spin period.
Systematic effects in the maps are small compared to either
the noise or the celestial signal:
the 95\% confidence upper limit 
for the pixel-pixel {\it rms} from all identified systematics is
less than 6 \muK ~in the worst channel.
A power spectrum analysis of the (A-B)/2 difference maps 
shows no evidence for additional undetected systematic effects.

\noindent
{\it Subject headings:} cosmic microwave background -- 
instrumentation: miscellaneous --
artificial satellites, space probes

\clearpage
\section{Introduction}
The large angular scale anisotropy of the cosmic microwave background (CMB) 
reflects the distribution of matter and energy in the early Universe,
before causal processes could operate to create the rich array of structures 
observed at the present epoch.  Precise observations of the CMB anisotropy
fix the initial conditions for different models of structure formation,
and have important implications for theories of the high-energy behavior
at the earliest times (e.g., inflation, cosmic defects).
The Differential Microwave Radiometers instrument 
on the {\it Cosmic Background Explorer} ({\it COBE}-DMR)
has completed its four-year observations of the microwave sky.
It has detected statistically significant signals
whose spatial morphology and frequency dependence 
are consistent with CMB anisotropy
(Bennett et al.\ 1996, 
Bennett et al.\ 1994,
Smoot et al.\ 1992).
Microwave emission from the Galaxy is weak at high latitudes
($|b| \gt 30\deg$)
and, with the notable exception of the quadrupole,
does not significantly contaminate the primordial CMB signal
(Kogut et al.\ 1996a,
Kogut et al.\ 1996b,
Bennett et al.\ 1992b).
The CMB is isotropic to high degree:
the detected anisotropy $\Delta T/T$ ranges from 
$10^{-6}$ to a few parts in $10^{-5}$
on angular scales larger than 7\deg.
At this level of sensitivity,
great care must be taken to demonstrate that the results
are not affected by microwave emission from
nearby objects hundreds to thousands of times hotter than the CMB
(e.g., the spacecraft, Earth, Moon, and Sun)
or from instrumental effects associated with 
changes in the orbital environment, telemetry, or data processing.

DMR is well characterized in terms of its response 
to various potential systematic effects.
Kogut et al.\ (1992) describe the analysis techniques and upper limits
on systematic artifacts in the first year sky maps.
An important result of their analysis was that,
owing to DMR's rapid scan pattern and good pixel-pixel connectedness,
most systematic effects couple only weakly to the sky maps.
Non-celestial signals from nearby objects or within the instrument
need not be removed to \muK ~precision in the time domain
for their effects in the map to be negligible.
Bennett et al.\ (1994) provide similar analysis and upper limits
for the two-year sky maps.

In this paper, we present the instrument calibration
and upper limits to systematic artifacts
in the final 4-year sky maps using data
from 22 December 1989 through 21 December 1993 UT.
Absolute calibration from external sources observed in flight
(the Moon and the Doppler dipole from the Earth's orbital motion about the Sun)
are in agreement with the more precise pre-launch calibration;
the calibration is determined to 0.7\% absolute 
with drifts smaller than 0.2\% per year.
We detect gain modulation 
$\Delta {\cal G}/{\cal G} \approx 10^{-5}$
at the orbit period 
during the two-month periods surrounding the June solstice, 
when the spacecraft flies through the Earth's shadow each orbit.  
The resulting modulation of the radiometric offset 
explains a previously detected instrumental signal
of unknown origin during these periods.

The largest uncertainties in the maps result from 
the instrument susceptibility to the Earth's magnetic field,
microwave emission from the Earth,
and upper limits to potential effects at the spacecraft spin period.
We correct the data for the magnetic response;
the resulting uncertainties are dominated by uncertainty in the 
magnetic field vector at each radiometer.
With the full four-year data set we detect emission from the Earth
when the Earth is 4\deg ~below the Sun/Earth shields or higher
(about 5\% of the data);
the detected signal is weak enough that correction is not required.
There is no evidence for additional systematic effects at the spin period,
but the precision with which we can rule out potential new effects
is limited by the instrument noise.
The quadrature sum of all systematic uncertainties in the 4-year maps,
after correction in the time domain,
yields upper limits
$\Delta T \lt 4 ~\muK$ {\it rms} over the full sky
in the 4 most sensitive channels.
Artifacts at this level would contribute less than 0.6\%
to the variance of the CMB signal at 7\deg ~resolution:
systematic effects do not limit the DMR maps.

\section{Instrument Description}
DMR consists of 6 differential microwave radiometers,
2 nearly independent channels (labeled A and B) 
at frequencies 31.5, 53, and 90 GHz
(wavelength 9.5, 5.7, and 3.3 mm).
Each radiometer measures the difference in power 
between two regions of sky separated by 60\deg,
using a heterodyne receiver switched at 100 Hz between
two corrugated horn antennas with beam width 7\deg 
~full width at half maximum
pointed 30\deg ~to either side of the spacecraft spin axis.
The 31A and 31B channels share a common antenna pair
in right and left circular polarizations;
the other four channels have independent antenna pairs
in a single linear polarization,
for a total of ten antennas.
The DMR antennas are mounted in 3 boxes spaced 120\deg ~apart
on the outside of the superfluid helium cryostat
containing the Far Infrared Absolute Spectrophotometer (FIRAS) and the
Diffuse Infrared Background Experiment (DIRBE).
The E-plane of linear polarization in each of the 53 and 90 GHz antennas
is directed radially outward from the spacecraft spin axis.
A shield surrounds the aperture plane to block radiation from the Earth and 
Sun.  The DMR antenna apertures lie approximately 6 cm below the shield plane.
The combined motions of the spacecraft spin (75 s period),
orbit (103 m period), and orbital precession ($\sim 1\deg$ per day)
allow each sky position to be compared to all others
through a highly redundant set of all possible difference measurements
spaced 60\deg ~apart.

The switched signal in each channel
undergoes RF amplification, detection, near-dc amplification,
and synchronous demodulation.
The demodulated signal is integrated for 0.5 s, digitized,
and stored in an on-board tape recorder.
A single local oscillator provides a common frequency reference
for the mixer in the A and B channels at each frequency.
The A and B channels also share a common enclosure and thermal regulation 
system but are otherwise independent.
We also sample the switched signal after one stage of dc amplification
but before synchronous demodulation;
the resulting ``total power'' signal serves as a low-precision
check on the system temperature and gain of the RF and first dc amplifier
with only minor contributions from celestial signals.

{\it COBE} was launched from Vandenberg Air Force Base
on 18 November 1989
into a 900 km, 99\deg ~inclination circular orbit
which precesses to follow the terminator.
Attitude control keeps the spacecraft 
pointed away from the Earth and
nearly perpendicular to the Sun.
Solar radiation never directly illuminates the aperture plane.
The Earth limb is below the shield for 95\% of the mission.
During the 2 months surrounding the June solstice,
the attitude control can not simultaneously block
both terrestrial and solar emission,
and the Earth limb rises above the shield 
as the spacecraft flies over the Arctic.
During the same 2 months, 
the spacecraft flies through the Earth's shadow over the Antarctic.
The resulting eclipse modulates spacecraft temperatures and voltages.

Several authors provide a more complete description of {\it COBE} and DMR.
Boggess et al.\ 1992 provide a mission overview.
Smoot et al.\ 1990 describe the DMR instrument.
Toral et al.\ 1990 and Wright et al.\ 1994 describe the DMR beam patterns.
Bennett et al.\ 1992a describe the instrument pre-flight calibration
and provide a schematic of the {\it COBE} aperture plane.

\section{Data Processing}
The digitized data from the on-board tape recorders 
are transmitted once per day to a ground station.
A software program merges the uncalibrated DMR differential data
with housekeeping (temperatures, currents, voltages, and relay states),
spacecraft attitude, and selected spacecraft archives
(magnetometers, momentum wheels, electromagnet currents).
The uncalibrated time-ordered data $S_{ij}(t)$ may be represented as
$$
S_{ij}(t) = \frac{1}{{\cal G}(t)} 
~[ ~T_i - T_j ~+ ~O(t) ~+ ~N(t) ~+ ~\sum_k Z_k(t) ~]
$$
where
$T_i$ and $T_j$ are the antenna temperatures 
of the two regions of sky observed at time $t$,
${\cal G}(t)$ is the gain factor providing calibration 
between antenna temperature and telemetry units,
$O(t)$ is the radiometric offset 
produced by small imbalances in the differential radiometer,
$N(t)$ is the random instrument noise,
and $Z_k(t)$ are the time-dependent systematic signals
from non-cosmological sources and instrumental effects.
A second program calculates the gain factor ${\cal G}^\prime(t)$,
the instrumental baseline $B(t)$, and corrections for known systematics
$W_n(t)$ to determine the calibrated differential temperatures
$$
D_{ij}(t) = {\cal G}^\prime(t)
[S_{ij}(t) ~- ~B(t)] ~- ~\sum_n W_n(t)
$$
which we refer to as the DMR time-ordered data set.

The instrument gain, beam patterns, and environmental susceptibilities
of each radiometer have been measured in pre-flight testing
(Smoot et al.\ 1990).
Based on these and on in-flight data, we reject data 
taken when the uncertainty in the systematic error model $W_n(t)$
becomes unacceptably large.
Table \ref{cut_list} 
lists cuts and corrections made in the 4-year DMR data processing.
We flag as unusable any datum for which the telemetry is unavailable
or of poor quality,
or which deviates from the mean by more than 5 standard deviations.
Such ``spikes'' are rarely noise outliers,
but result instead from 
observations of the Moon
or instrument changes such as 
the end of the noise source calibration sequence.
The exponential decay of the noise source power
contaminates the first few samples
after the noise sources have been commanded off;
as a result, we conservatively discard the entire 32 s frame
after each in-flight calibration.

We flag as unusable for sky maps
any datum taken with the Moon within 21\deg ~of an antenna beam center,
although we do use these data for lunar calibration and to map the
beam pattern in flight.
All other data are corrected for a model of lunar emission.
We reject data taken when the Earth limb is 1\deg ~below the shield
or higher (3\deg ~for the 31.5 GHz channels) but do not otherwise correct
for Earth emission.
We correct data taken with Mars, Jupiter, or Saturn within 15\deg
~of an antenna beam center but do not reject these data as unusable.
We correct the time-ordered data for 
the instrument response to the Earth's magnetic field,
the 3.2\% correlation between successive observations
caused by the lock-in amplifier low-pass filter,
and for the Doppler dipole from the satellite motion about the Earth 
and the Earth's orbital motion about the solar system barycenter.
An orbitally-modulated signal is present
during the 2-month ``eclipse season'' surrounding the June solstice,
related to thermally-induced gain variations in the amplification chain.
An empirical model based on tracers with high signal to noise ratio
removes approximately 2/3 of this signal.
Estimated residuals, after correction, 
are small compared to the increase in data and sky coverage in the 53 and 90 
GHz channels (recall that orbitally modulated signals couple only weakly to the 
sky maps, so that a small systematic signal accompanied by a larger increase in 
sensitivity can be a worthwhile trade-off).
The amplitude of the effect, and hence the residual uncertainty,
is larger in the 31A and 31B channels;
we reject data taken during the period
May 1 through August 4 of each year for the 31A channel
and 
May 21 through July 24 for the 31B channel.
The 31B channel suffered a permanent increase in receiver noise
on 4 October 1991.
We flag as unusable data when the instrument noise is unstable,
and use the noise during stable periods to weight all analyses.
As a result, the 31B results are approximately 30\% noisier
than the 31A channel.

We fit the time-ordered data to a pixelized sky map and 
systematic terms by minimizing the $\chi^2$ sum
\begin{equation}
\chi^2 = \sum_t 
~[ 
~\frac{ D_{ij}(t) ~- ~(T_i - T_j) }{\sigma_{ij}(t)}
~]^2
\label{chisq_eq}
\end{equation}
where $\sigma_{ij}(t)$ is the instrument noise
(Bennett et al.\ 1996).
A sparse-matrix algorithm performs the inversion
and yields a set of pixel temperatures $T_i$ and
coupling coefficients $a_n$ to specified systematic effects
(Janssen \& Gulkis 1992, 
Jackson et al.\ 1992, 
Torres et al.\ 1989).

Several methods exist to derive limits to systematic signals $Z_k(t)$
in the time-ordered data and their projection $\Delta T_{\rm sys}$
in the sky maps.
Given a time-dependent model $W_n(t)$,
a linear least-squares method solves simultaneously
for the systematic coupling coefficient $a_n$ in the time domain
and the sky temperatures $T_i$ in the map
(Eq.\ \ref{chisq_eq}). 
This procedure automatically removes the best estimate of the particular
systematic signal from the sky map.
We make a map $\Delta T_{\rm sys}$ of the removed systematic effect by 
subtracting two solutions to Eq.\ \ref{chisq_eq} 
which differ only in the presence or 
absence of the model function $W_n(t)$.  
The residual uncertainty $\delta T_{\rm sys}$ after correction
is determined by multiplying the systematic map $\Delta T_{\rm sys}$
by the fractional uncertainty in the coupling coefficient:
\begin{equation}
\delta T_{\rm sys} = \frac{\delta a_n}{a_n} \Delta T_{\rm sys} .
\label{syserr_map_eq}
\end{equation}

In some cases,
the model function $W_n(t)$ may not be well specified {\it a priori}.
We may re-bin the time-ordered data into
a coordinate system where the signal will add coherently
to obtain limits to both
the shape and amplitude of the systematic function $Z_k(t)$.
We then use the binned data and associated uncertainties,
sampled according to the DMR observation pattern,
as the model function $W_n(t)$ in Eq.\ \ref{chisq_eq}
to derive the systematic error map and uncertainty.
For example, microwave emission from the Earth diffracted over the shield
has a complicated time dependence whose exact form depends sensitively
upon the relative geometry of the Earth, shield, and horn antennas.
We bin the data by the location of the Earth relative to the {\it COBE}
aperture plane and use the binned data 
both to evaluate various models $W_n$ of the diffraction
and to derive model-independent
limits to Earth emission in the DMR maps.
We also bin the time-ordered data by the spacecraft orbit and spin angles
to search for signals at those periods.

Several tests limit systematic effects
without requiring any {\it a priori} information.
Fourier transforms of the calibrated $D_{ij}(t)$
provide a powerful tool to limit potential effects.
Common-mode signals, particularly celestial emission, cancel
in the (A-B)/2 ``difference map'' linear combination of the
A and B channels at each frequency,
or in similar difference maps 
from different time ranges in a single channel.
Analysis of the difference maps
provides model-independent limits to the combined effects of all
systematic effects which are not identical 
in different channels or at different times.

\begin{figure}[t]
\epsfxsize=6.0truein
\epsfbox{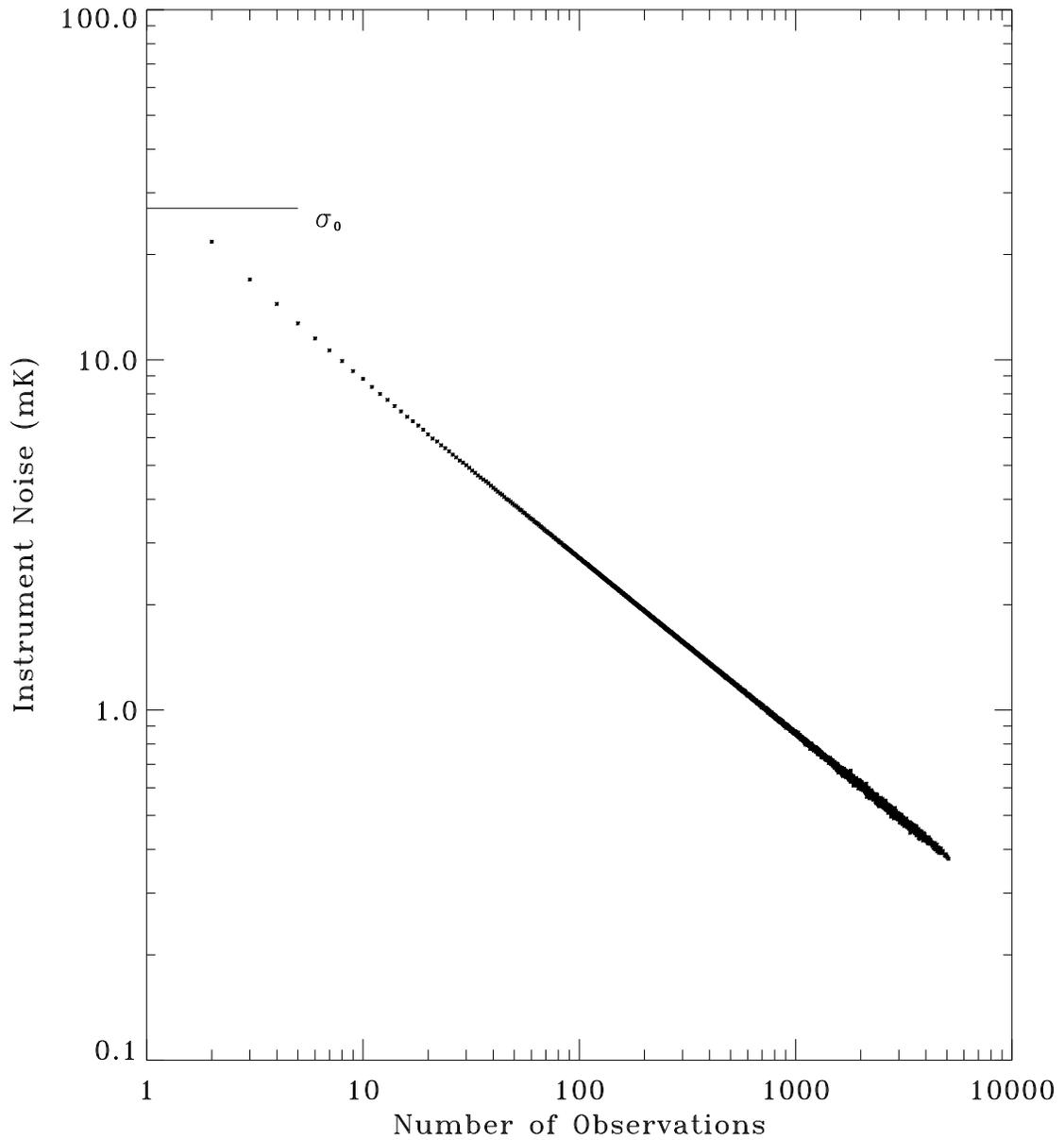}
\caption{
Instrument noise for the 53B radiometer, sorted by integration time.
The time-ordered data are first sorted by the orientation of the two antennas,
and all observations with the spacecraft in a fixed position in inertial 
coordinates are combined.  
Temperature patterns fixed on the sky affect the mean temperature difference
of each such pixel combination but do not affect the standard deviation.
The plot shows the uncertainty in the mean values (``instrument noise'').
There is no evidence for a noise floor caused by non-celestial signals.
}
\label{rms_vs_nobs_fig} 
\end{figure}

\begin{figure}[t]
\epsfxsize=6.0truein
\epsfbox{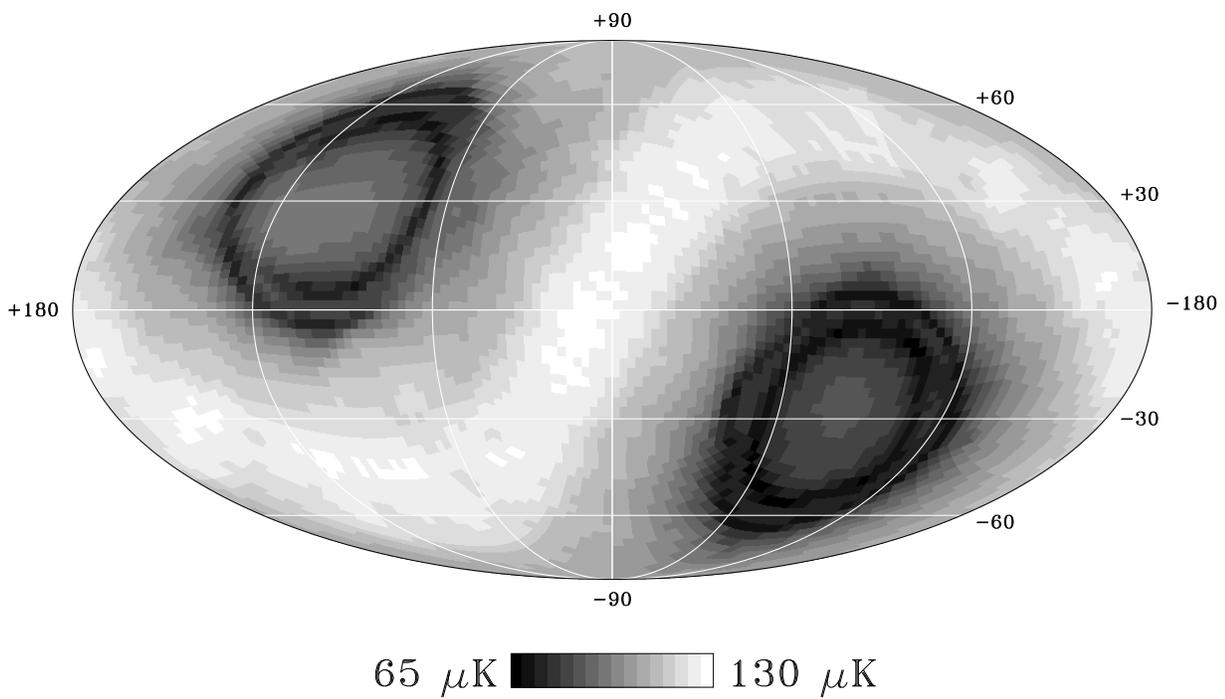}
\caption{
Noise pattern of the 53B channel for 4 years of data.
The map is a Mollweide projection in Galactic coordinates.
The lunar cut causes the striping in the ecliptic plane,
while the cut for the Earth limb causes the north-south asymmetry
near the ecliptic poles.
}
\label{nobs_fig}
\end{figure}

The instrument noise is well described 
by a Gaussian probability distribution
with standard deviation depending on 
the number of discrete observations $N_i$ per pixel,
$$
n_i = \frac{\sigma_0}{\sqrt{N_i}},
$$
where $\sigma_o$ is the {\it rms} noise per 0.5 s observation
(Bennett et al.\ 1996).
Figure \ref{rms_vs_nobs_fig} shows the instrument noise
sorted by the number of observations linking two pixels
at 60\deg ~separation.  
There is no evidence for a ``noise floor'' from effects
which do not average out with time.
Figure \ref{nobs_fig} shows the observation pattern in the 53B channel
for the 4-year mission after applying all cuts.
Correlations between pixels caused by the 60\deg ~antenna separation
are negligible (Lineweaver et al.\ 1994).
Because {\it COBE} is in a polar orbit, regions approximately
30\deg ~from the celestial poles are observed each orbit,
while regions near the celestial equator
are only observed for 4 months of each year.
The lunar cut further reduces usable data near the ecliptic plane.
Since the model functions $W_n(t)$ are typically not well described
by a set of temperatures $T_i$ fixed on the sky,
gradients in the sky coverage are apparent 
in many of the systematic error maps.

\section{Calibration}
The primary DMR calibration consists of the 
radiometric comparison, before launch,
of cold ($\sim$77 K) and warm ($\sim$300 K) 
full-aperture blackbody targets.
In addition, noise sources inject $\sim$2 K of broad-band power
into the front end of each radiometer between the horn antenna
and the Dicke switch.
Near-simultaneous observations of the blackbody targets and the noise sources 
calibrate the antenna temperature of each noise source
and permit the transfer of the blackbody calibration standard
to the flight observations (Bennett et al.\ 1992a).
Observations of the Moon and CMB dipole provide additional in-flight
calibration.
Table \ref{cal_summary} summarizes the DMR calibration over the 4-year mission.

\subsection{Absolute Calibration}
\begin{figure}[t]
\epsfxsize=6.0truein
\epsfbox{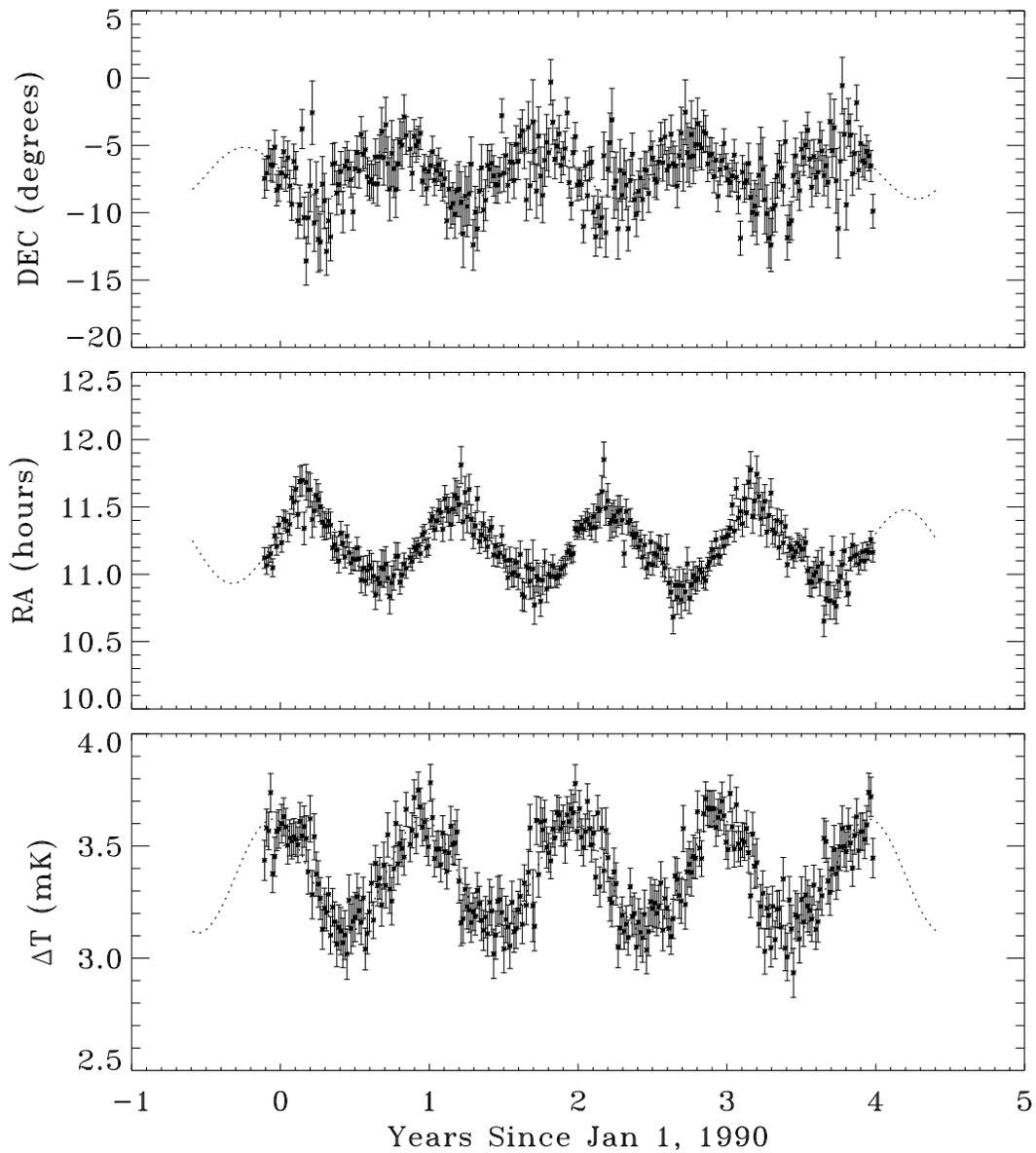}
\caption{
Modulation of the CMB dipole resulting from the 
Doppler effect of the Earth's orbital motion about the Sun
(channel 53B).  
Each datum represents 5 days.
The amplitude of the modulation provides an independent absolute calibration.
}
\label{dip_vs_time_fig}
\end{figure}

The pre-flight absolute calibration has been adjusted 
for two effects observed in the first year of the mission:
the 31B gain was decreased by 4.9\%
and the 90B gain increased by 1.4\%
(Kogut et al.\ 1992).
Errors in the absolute calibration 
create systematic artifacts in two ways:
errors in the amplitude of detected structures in the sky,
and artifacts in maps from which a model of celestial emission
(e.g., lunar emission) have been removed.
Observations of the Doppler dipole from the Earth's orbital velocity
provide an independent absolute calibration in flight.
Motion with velocity ${\bf \beta}$ through an isotropic blackbody radiation
field at temperature $T_0$ 
creates an observed thermodynamic temperature distribution
\begin{equation}
T(\theta) = 
\frac{T_0 (1-\beta^2)^{1/2}}{1 - \beta \cos(\theta)}
\approx
T_0 [ 1 + \beta \cos(\theta) + O(\beta^{2}) ]
\label{dipole_eq}
\end{equation}
(Peebles \& Wilkinson 1968).
The CMB spectrum is well described by a blackbody 
with $T_0$ = 2.728 K (Fixsen et al.\ 1996).
The dipole caused by the Earth's orbital velocity 
about the solar system barycenter
(29.27 km s$^{-1}$ to 30.27 km s$^{-1}$)
provides a calibration signal
$\Delta T \approx 270$ \muK 
~whose known spatial and temporal dependence allows simple separation
from other astrophysical signals.
Figure \ref{dip_vs_time_fig} shows the modulation of the CMB dipole
observed in the 53B channel throughout the 4-year DMR mission.
We fit the time-ordered data for a fixed sky map 
plus a Doppler calibration term,
including the change in the orbital speed 
from the eccentricity of the Earth's orbit.
Table \ref{abs_cal} compares the Doppler calibration to the primary calibration.
Values larger than unity indicate a channel in which the observed Doppler 
dipole is larger than predicted.
The Doppler calibration is in agreement with the more precise
pre-launch absolute calibration
and shows no evidence for calibration shifts at the few percent level.

If we accept the pre-launch calibration as accurate,
the Earth Doppler dipole provides a determination
of the absolute CMB temperature $T_0$.
We may use 
Eq.\ \ref{dipole_eq}
and the Doppler calibration factors from Table \ref{abs_cal} 
to infer the CMB absolute temperature:
$T_0 ~= 2.83 \pm 0.07$ K at 31.5 GHz,
$2.71 \pm 0.03$ K at 53 GHz,
and $2.72 \pm 0.04$ K at 90 GHz
for a weighted mean $T_0 = 2.725 \pm 0.020$ K (68\% CL),
in excellent agreement with the FIRAS spectrum.

\begin{figure}[t]
\epsfxsize=6.0truein
\epsfbox{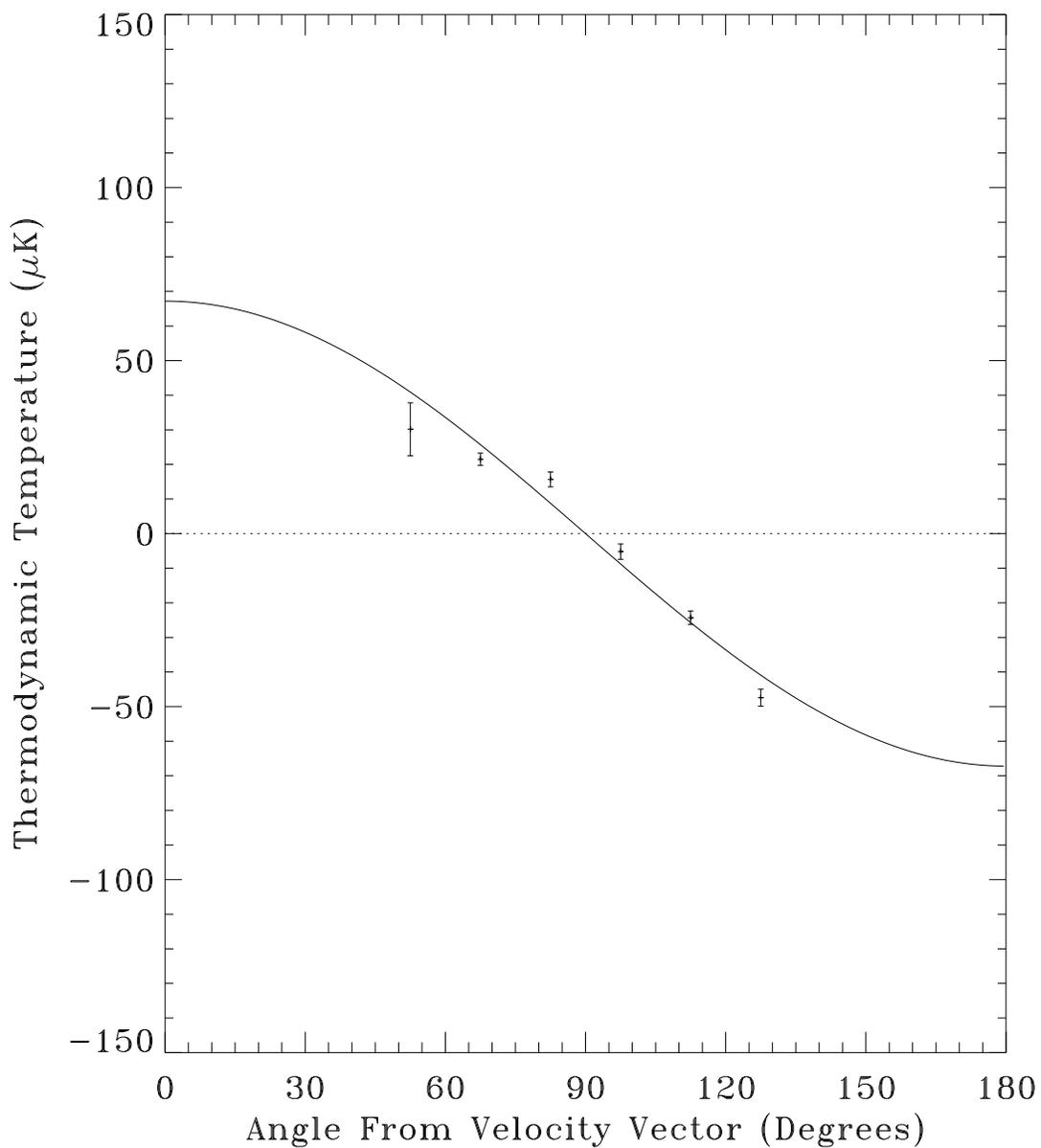}
\caption{
53 GHz (A+B)/2 summed data mapped in a coordinate system 
fixed with respect to the satellite orbital velocity vector.
The data have been binned by angle relative to the {\it COBE}
orbital velocity vector and corrected for the 0\ddeg5 smoothing
of the DMR beam.  The attitude control system prevents full coverage
in this coordinate system.  
The Doppler effect provides a known signal 
with amplitude $\Delta T/T \sim 10^{-5}$.
}
\label{COBE_vel_fig}
\end{figure}

Figure \ref{COBE_vel_fig} shows the Doppler dipole resulting from 
{\it COBE}'s 
7.5 km s$^{-1}$ orbital velocity about the Earth.
We have mapped the time-ordered data in a coordinate system
co-moving with the satellite velocity vector,
and binned the resulting map temperatures by the angle from the 
velocity vector.
The attitude control system prevents DMR from observing all angles
in this coordinate system.
Within the range of angles observed, we find good agreement with the
predicted 70 \muK ~dipole.
The {\it COBE} velocity dipole demonstrates
the sensitivity to signals $\Delta T/T \sim 10^{-5}$.

The Moon provides a second external absolute calibration source.
DMR observes the Moon in the antenna main beam
for $\sim$6 days each at both first and third quarters
(Bennett et al.\ 1992a).
The antenna temperature of the Moon within integration time $\tau$ 
is given by
$$
T_{A,{\rm Moon}} ~= ~\frac{1}{\tau} 
~\int^{t + \tau}_{t} dt
~\int d\Omega ~T_{\rm Moon} ~P[\theta(t), \phi(t)]
~/ 
~\int d\Omega ~P(\theta,\phi)
$$
where
$P(\theta,\phi)$ is the DMR beam pattern.
We integrate the physical temperature and microwave emission properties
across the lunar disk to estimate $T_{\rm Moon}$ as a function of
lunar phase and Sun-Moon distance
(Keihm 1982,
Keihm \& Gary 1979,
Keihm \& Langseth 1975).
The gain factors derived from observations of the Moon
show a pronounced dependence on lunar phase,
with peak-to-peak variation of 3.7\%, 5.3\%, and 6.2\%
at 31.5, 53, and 90 GHz, respectively.
Longer term analysis shows an additional annual modulation
with peak-to-peak amplitude 2\%.
These apparent gain modulations are not present
in any method which does not involve lunar observations
(i.e., on-board noise sources or the CMB dipole).
We thus ascribe them to real time-dependent effects in $T_{\rm Moon}$
which are not duplicated in the Keihm model.
See Jackson et al.\ 1996 for further analysis of $T_{\rm Moon}$.

The lunar calibration modulation is repeatable over the 4-year mission
and may be removed empirically to provide a stable
``standard candle'' for relative gain analysis.
We adopt the combined peak-peak modulation
as an estimate of the uncertainty of the lunar model
for the absolute calibration.
Table \ref{abs_cal} shows the mean lunar calibration compared to the
noise source calibration.
The channel averaged lunar absolute calibration is 1.8\% larger than 
the pre-launch absolute calibration,
well within the systematic limitations of the lunar model.

The Moon also serves to cross-calibrate the A and B channels.
Systematic uncertainties in the model for $T_{A,{\rm Moon}}$
cancel in the A/B ratio at each frequency.
The resulting ratio (Table \ref{cross_cal})
places a limit to how well celestial emission will be expected to
cancel in the (A-B)/2 ``difference maps.''
A similar analysis for the Earth velocity Doppler effect
is consistent within much larger uncertainties.
The A and B channels are cross-calibrated within 0.4\%,
well within the  accuracy of the pre-flight calibration.

\subsection{Relative Calibration}
\begin{figure}[t]
\epsfxsize=6.0truein
\epsfbox{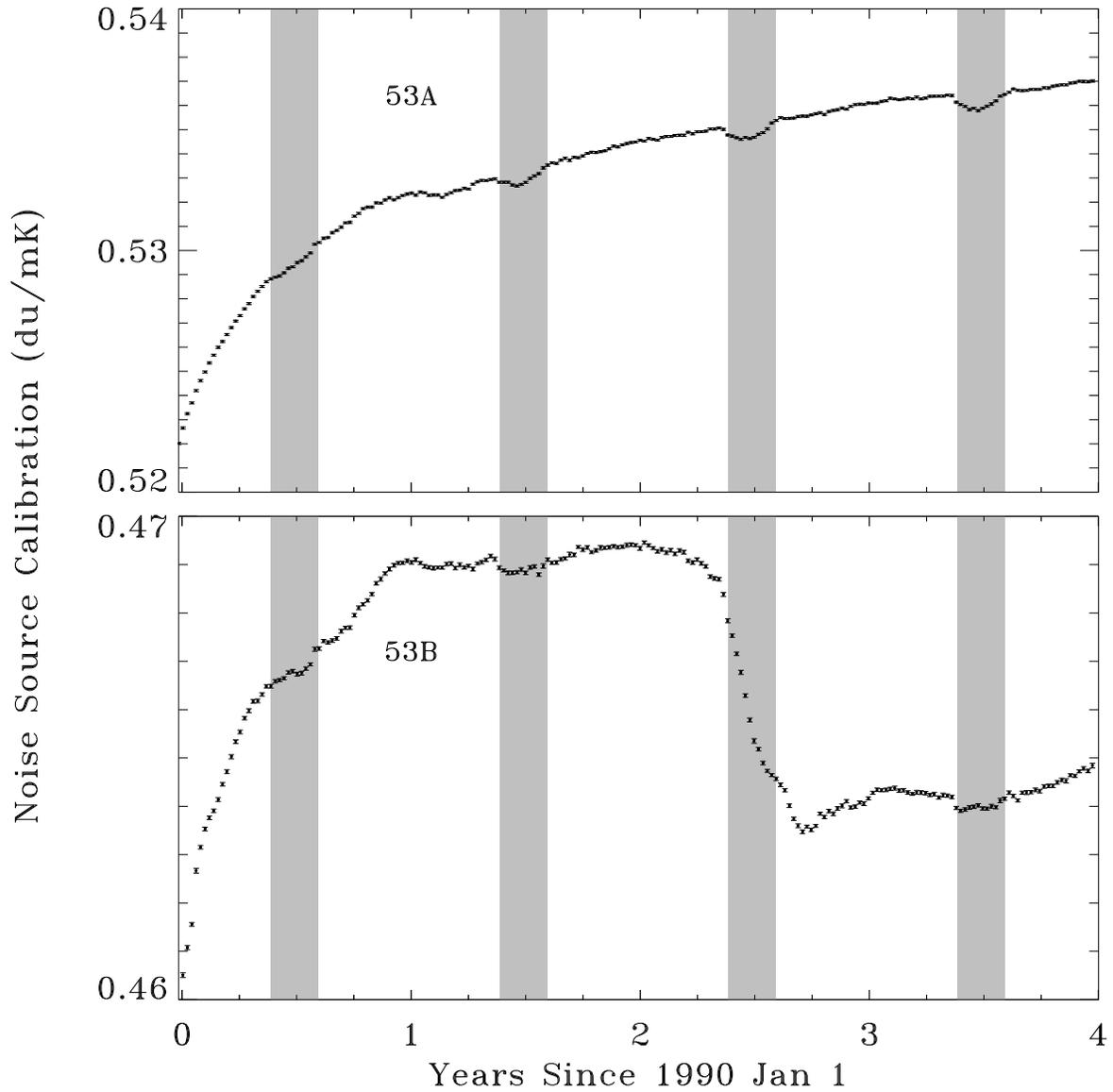}
\caption{
Calibration factor ${\cal G}^\prime (t)$ derived from on-board noise sources
for the 53A and 53B channels.  Each datum represents one week.
The gray bands show the ``eclipse season'' surrounding the June solstice.
The gain is stable to better than 3\% throughout the 4-year mission.
}
\label{gain_vs_time_fig}
\end{figure}

The noise sources are commanded on for 128 seconds every two hours.
Provided the power broadcast by each device is constant in time,
they provide a standard to monitor time-dependent
changes in the instrument calibration.
Figure \ref{gain_vs_time_fig} shows the calibration ${\cal G}^\prime (t)$
derived from the noise sources
for the 53A and 53B channels over the full 4-year mission.
The gain is stable to better than 3\% throughout the 4-year mission.
Gain drifts {\it per se} do not create systematic artifacts 
provided the noise sources track the true calibration ${\cal G}(t)$.

Each noise source is observed in both the A and B channels.
The ratio of the two noise sources in a single channel
provides information on the noise source stability,
since the instrument calibration cancels.
The ratio of a single noise source observed in two channels
provides information on gain stability,
since the noise source performance cancels.
Based on these ratios, we correct the data for 3
step changes in noise source broadcast power:
a 0.69\% increase in power for the 90 GHz ``down'' noise source 
	on 17 March 1990,
a 0.69\% increase in power for the 31 GHz ``down'' noise source 
	on 11 February 1992,
and
a 0.34\% increase in power for the 90 GHz ``down'' noise source 
	on 26 November 1993.
An additional anomaly occurred on 1 September 1993 in the 90 GHz data,
when both noise sources showed step changes in a pattern inconsistent
with a simple change in power or instrument calibration.
A 0.29\% step increase occurred for the 90A channel ``up'' noise source
which was not mirrored by the same noise source observed in the 90B channel.
At the same time, the ``down'' noise source in the 90B channel
decreased in power by 0.66\%, 
unaccompanied by a similar change in the 90A channel.
Since a calibration change would show up 
as a step in both the ``up'' and ``down'' noise source signals, 
while a noise source power change would show up
in both the 90A and 90B channels,
we can rule out simple models involving a single component.
We currently have no explanation for this event,
but remove its effects in software
by increasing the ``up'' noise source by 0.29\% for the 90A channel only
and decreasing the ``down'' noise source by 0.66\% for the 90B channel only.

\subsubsection{Long-Term Calibration Drifts}
\begin{figure}[t]
\epsfxsize=6.0truein
\epsfbox{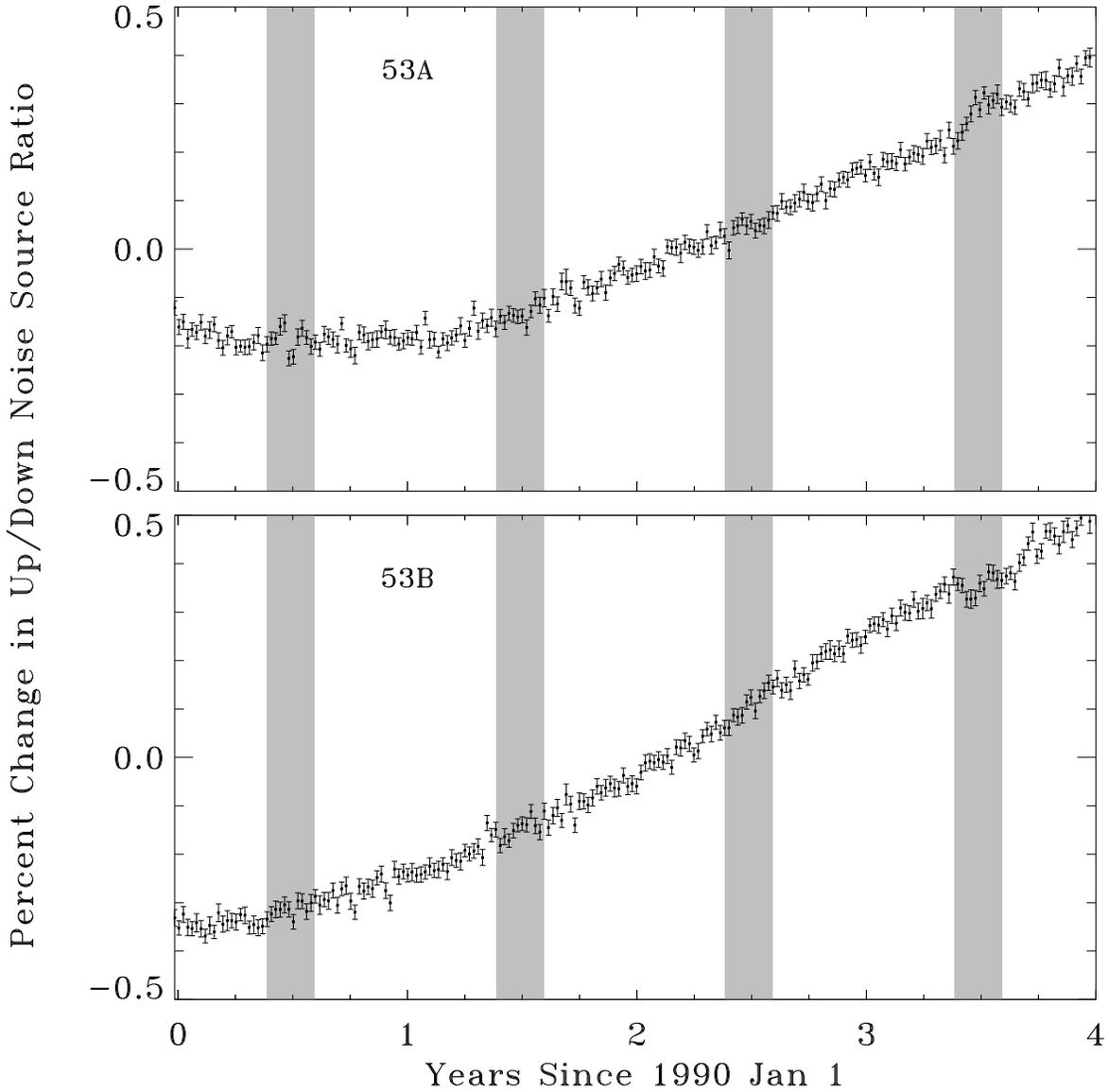}
\caption{
Ratio of ``up'' and ``down'' noise sources vs time for 53A and 53B channels.
Each datum represents one week.
The gray bands show the ``eclipse season'' surrounding the June solstice.
Uncorrected linear drifts are smaller than 0.2\% per year at 95\% confidence.
}
\label{ns_ratio_fig}
\end{figure}

The gain solutions in the 90A and 90B channels are corrected for
linear drifts of 0.81\% yr$^{-1}$ and 0.87\% yr$^{-1}$ respectively
(Kogut et al.\ 1992, Bennett et al.\ 1994).
We place limits on un-corrected long-term drifts in noise source power by
examining the ratio of the A and B noise sources in each channel.
The instrument calibration cancels in this ratio,
which places a lower limit to the long-term accuracy 
of the calibration solution
(which assumes that the noise sources do not change in time).
Figure \ref{ns_ratio_fig} shows this ratio for both the 53A and 53B channels.
Channel 53A shows a stable ratio for the first year 
followed by a linear drift, 
while channel 53B shows an approximately linear drift throughout the mission.
The fact that the ratio of the same physical devices 
does not have the same shape in both channels 
indicates that the observed changes
are caused by a process more complicated than a simple change in
noise source broadcast power.
Table \ref{linear_drifts} shows the limits to
linear drifts in all 6 channels based on the noise source ratios.

The CMB dipole provides a continuously observed signal of 3 mK amplitude,
which we use to limit long-term errors in the noise source calibration.
We fit the time-ordered data to the form
$$
\Delta T ~= ~(1 + bt) ~\sum_{\ell = 1}^{2} \sum_{m = -\ell}^\ell
~a_{\ell m} Y_{\ell m},
$$
i.e., a spatially fixed dipole and quadrupole 
whose amplitudes change linearly with time.  
Since the CMB is effectively a constant 
for the four years of DMR observations,
we can interpret the fitted parameter $b$ in terms of 
a linear drift in the true calibration ${\cal G}(t)$
relative to the noise source solution ${\cal G}^\prime (t)$.
Table \ref{linear_drifts} shows the resulting limits to calibration drifts.
There are no drifts significant at 95\% confidence level.

The Moon also serves as a standard candle 
once the annual and phase variations are empirically removed.  
We have analyzed the ratio of the corrected lunar calibration
to the noise source calibration to search for long-term
drifts in the noise source solution (Table \ref{linear_drifts}).
The drifts inferred from the noise sources, dipole, and Moon
are in general agreement and provide some evidence that calibration drifts are 
dominated by changes in the power emitted by the noise sources.
The significance of the coefficients is difficult to evaluate.
The uncertainties in Table \ref{linear_drifts} are statistical only.
Since the lunar results are possibly affected by long-term artifacts
related to the annual and phase variations, 
and since the noise source ratios 
are clearly more complicated than a simple linear drift,
we do not use a weighted estimate of the three techniques.
Instead, we adopt the unweighted mean and use the scatter among the three 
techniques as an estimate of the uncertainty.
Table \ref{cal_summary} shows the resulting limits on
long-term calibration drifts in the 4-year DMR data set.  

\subsubsection{Orbital Calibration Drifts}
\begin{figure}[t]
\epsfxsize=6.0truein
\epsfbox{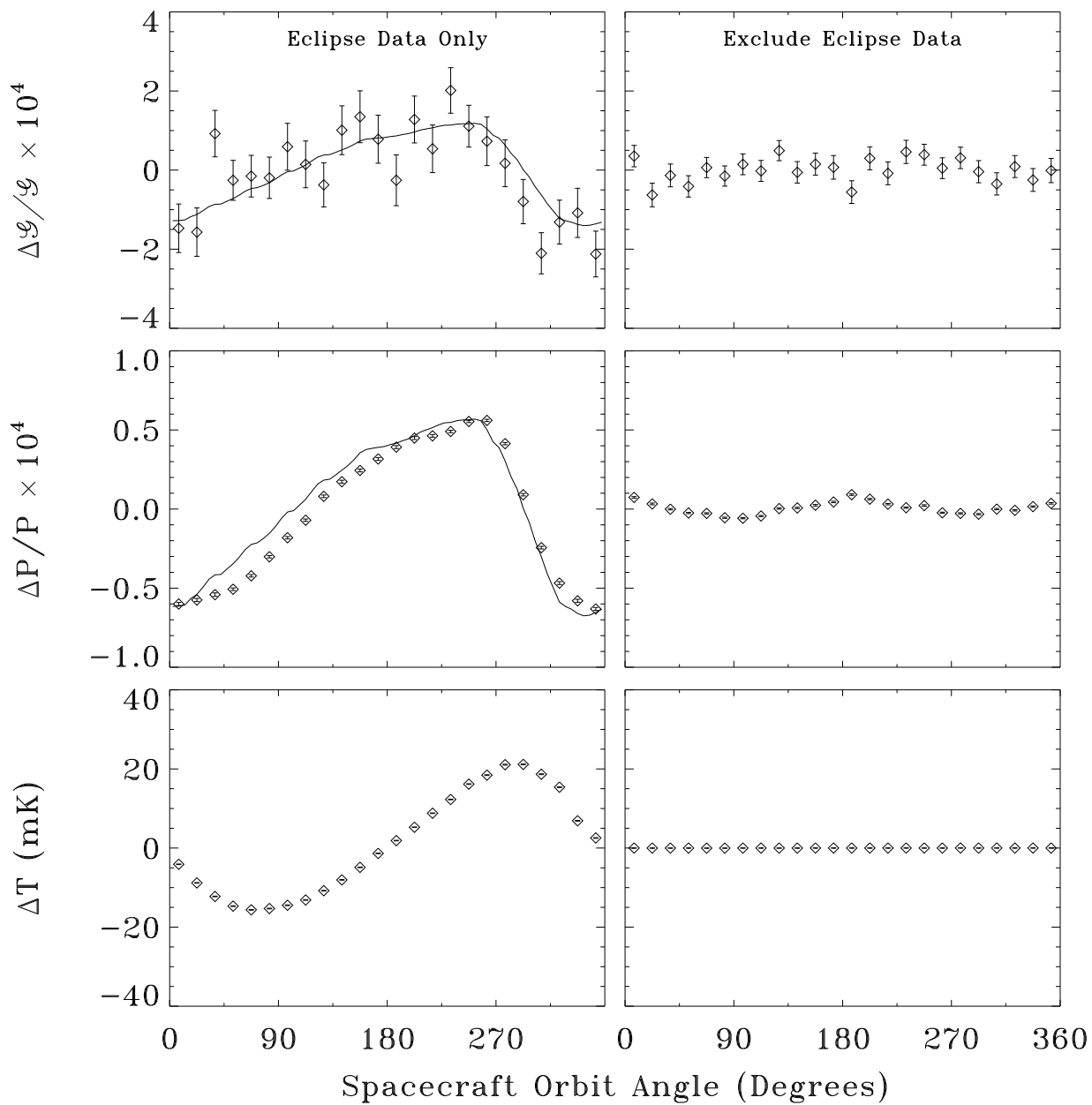}
\caption{
Calibration signals binned at the orbit period
for eclipse data (left panels) and non-eclipse data (right panels).
(top) Noise source calibration residuals.
(middle) Total power residuals.
(bottom) Lock-in amplifier temperature.
The noise source and total power plots during eclipse season
are overlaid with the scaled IPDU thermistor temperature
to demonstrate the similarity of the wave forms (see text).
}
\label{cal_vs_orbit_fig}
\end{figure}

The noise sources provide direct calibration information every two hours.  
We interpolate the noise source gain solution ${\cal G}^\prime (t)$
using 48 hours of data fitted to a cubic spline with one interior knot.
The resulting gain solutions are smooth at the daily boundaries
but can not respond to gain variations 
with periods shorter than about 8 hours.
We estimate gain variations at the orbit period by
fitting a long-term baseline to the noise source calibrations
and binning the residuals by the spacecraft orbit angle 
relative to the ascending node.
Long-term plots of the calibration (Fig. \ref{gain_vs_time_fig})
show effects related to the ``eclipse season'' 
surrounding the June solstice.
Figure \ref{cal_vs_orbit_fig} shows the binned calibration residuals
during eclipse season 
and for ``non-eclipse'' data (the rest of each year).
The noise sources (top panels) show gain variation 
with amplitude $\Delta {\cal G}/{\cal G} ~\approx ~1.6 \times 10^{-4}$
during the eclipse season.
The shape of the gain variation 
is nearly identical to the spacecraft temperature variations,
here (superimposed solid line) represented by a thermistor in the 
Instrument Power Distribution Unit (IPDU).
Since the IPDU is not thermally controlled,
it shows larger temperature variations
which are less affected by the housekeeping digitization.
The total power (middle panels), binned in a similar fashion,
shows similar modulation,
supporting the existence of a real 
orbitally-modulated gain variation during the eclipse season.
The amplitude of the total power variation,
$\Delta P/P ~\approx ~6.1 \times 10^{-5}$,
is smaller than the noise source variation,
as expected
since the total power does not sample the entire amplification chain.

The mixer/preamp assembly and lock-in amplifier
are maintained in a thermally controlled box.
Figure \ref{cal_vs_orbit_fig} also shows the lock-in amplifier temperature
(bottom panels) during eclipse and non-eclipse data.
Thermal variations of $\sim$10 mK amplitude at the orbit period
are observed during eclipse season, 
compared to $\sim$6 K for the unregulated IPDU.
The thermal susceptibility of the amplifiers,
measured prior to launch,
is $\Delta {\cal G}/{\cal G} \approx 1\% ~{\rm K}^{-1}$
(Bennett et al.\ 1992a).
The observed orbital gain variations are consistent 
with thermal modulation at the 10 mK level.

The existence of gain modulation $\Delta {\cal G}/{\cal G} \approx 10^{-4}$
with wave form similar to the IPDU thermistor
provides a plausible explanation for 
a previously detected signal during eclipse season
({\S}5.2; see also Kogut et al.\ 1992 and Bennett et al.\ 1994).
An empirical fit to the IPDU thermistor and voltage monitors
removes most of this signal;
we do not explicitly correct the noise source solutions
to model the orbital gain variation during eclipse season.
Outside of eclipse season, the orbital environment is stable.
The amplifier temperatures and direct noise source calibration 
show no modulation at the orbit period.
The total power shows slight orbital modulation
linked to variations in the input celestial signal 
from the CMB dipole and Galactic plane.
There is no evidence for orbitally modulated gain variation
outside of the eclipse season
at the level of a few parts in $10^5$
(Table \ref{cal_summary}, column 6).

The noise source calibrations require 128 s, longer than the
73 s spin period.
We obtain limits on calibration modulation at the spin period
by binning the total power signal (sampled every 8 s)
and scaling the resulting limits using the ratio of total power 
to gain variations observed at the orbit period,
$\Delta {\cal G}/{\cal G} = 3.1 \Delta P/P$.
We find no variation in either the total power or the amplifier temperature
at the spin period (Table \ref{cal_summary}, column 7),
with limits $\Delta {\cal G}/{\cal G} ~\lt ~2 \times 10^{-6}$
(95\% CL).

Gain modulation can create systematic artifacts in two ways: 
through observations of the sky 
at different relative calibration (``striping''),
and by modulating the radiometric offset.
The relative importance of the two effects depends on the time scale.
Over long periods, the offset signal $\Delta {\cal G}/{\cal G} \times O(t)$
is removed by the fitted baseline, leaving striping as the primary gain 
artifact.  
On time scales comparable to the spin or orbit periods,
offset modulation becomes dominant.
Since the offsets can approach 1 K,
a gain modulation as small as $10^{-4}$
can create a 100 \muK ~signal.
We derive upper limits on the systematic artifacts in the DMR sky maps
resulting from calibration drifts
at the spin, orbit, or longer periods
by adding terms to Eq.\ \ref{chisq_eq} 
of the form
$W_n = \Delta {\cal G} (S_{ij}(t) + B(t))$
where $\Delta {\cal G}(t)$ is taken 
from the 95\% confidence upper limits in Table \ref{cal_summary}.
That is, we simulate the effect of an uncorrected calibration modulation
in the striping of the sky,
the modulation of the instrumental offset,
and possible cross-talk with other systematic effects.
Systematic artifacts
from residual calibration drifts and modulation
are negligible,
creating {\it rms} variations 0.4 \muK
~or smaller (95\% CL) in the most sensitive sky maps.

\section{Environmental Effects}
The {\it COBE} orbit provides a generally benign environment 
for the DMR instrument.
The radiometers are above the Earth's atmosphere,
and the terminator-following orbit prevents large temperature changes.
Kogut et al.\ (1992) review various effects associated with the orbital 
environment.
The largest signals result from the
response of the Dicke switch to the Earth's magnetic field,
and the thermal response of the amplifiers
to temperature changes when the orbit passes into the Earth's shadow.
We correct for both of these effects; residual artifacts 
in the sky maps are small.

\subsection{Magnetic Susceptibility}
The amplification chain in each channel
is connected to the antennas
using a latching ferrite circulator
switched at 100 Hz by an applied magnetic field.
An external magnetic field 
(from the Earth or the 
electromagnets used to control the spacecraft angular momentum)
modulates the insertion loss of the switch
and creates a time-dependent signal
described by the vector coupling
\begin{equation}
Z_{\rm magnetic}(t) = 
\vec{\beta} ~\cdot ~\vec{B}(t) 
\label{magsus_eq}
\end{equation}
where $\vec{B}(t)$
is the magnetic field vector.
We express the magnetic susceptibility vector 
$\vec{\beta}$ 
in an orthonormal coordinate system fixed with respect to the spacecraft:
$\beta_X$ is the susceptibility along the X axis
(antiparallel to the spin axis),
$\beta_R$ is the susceptibility along the radial axis
(directed outward between two antennas),
and
$\beta_T$ is the susceptibility along the transverse axis
(from the positive horn to the negative horn).
The antennas are pointed 30\deg ~to either side of the $X$ axis;
magnetic signals from the $\beta_X$ susceptibility are not spin modulated
and produce a signal at the orbit period only.
Both the $R$ and $T$ signals are spin modulated.
The $\beta_R$ susceptibility produces an apparent temperature gradient
oriented across the magnetic field (east-west) 
but at right angles to the antenna pointing.
The $\beta_T$ susceptibility produces an apparent temperature gradient
oriented along the magnetic field (north-south)
in phase with the antenna pointing.
The inclination of the {\it COBE} orbit with respect to the magnetic poles
breaks the degeneracy which would otherwise 
make the $\beta_T$ susceptibility indistinguishable 
from a celestial dipole 
aligned with the Earth's magnetic field.

\begin{figure}[t]
\epsfxsize=6.0truein
\epsfbox{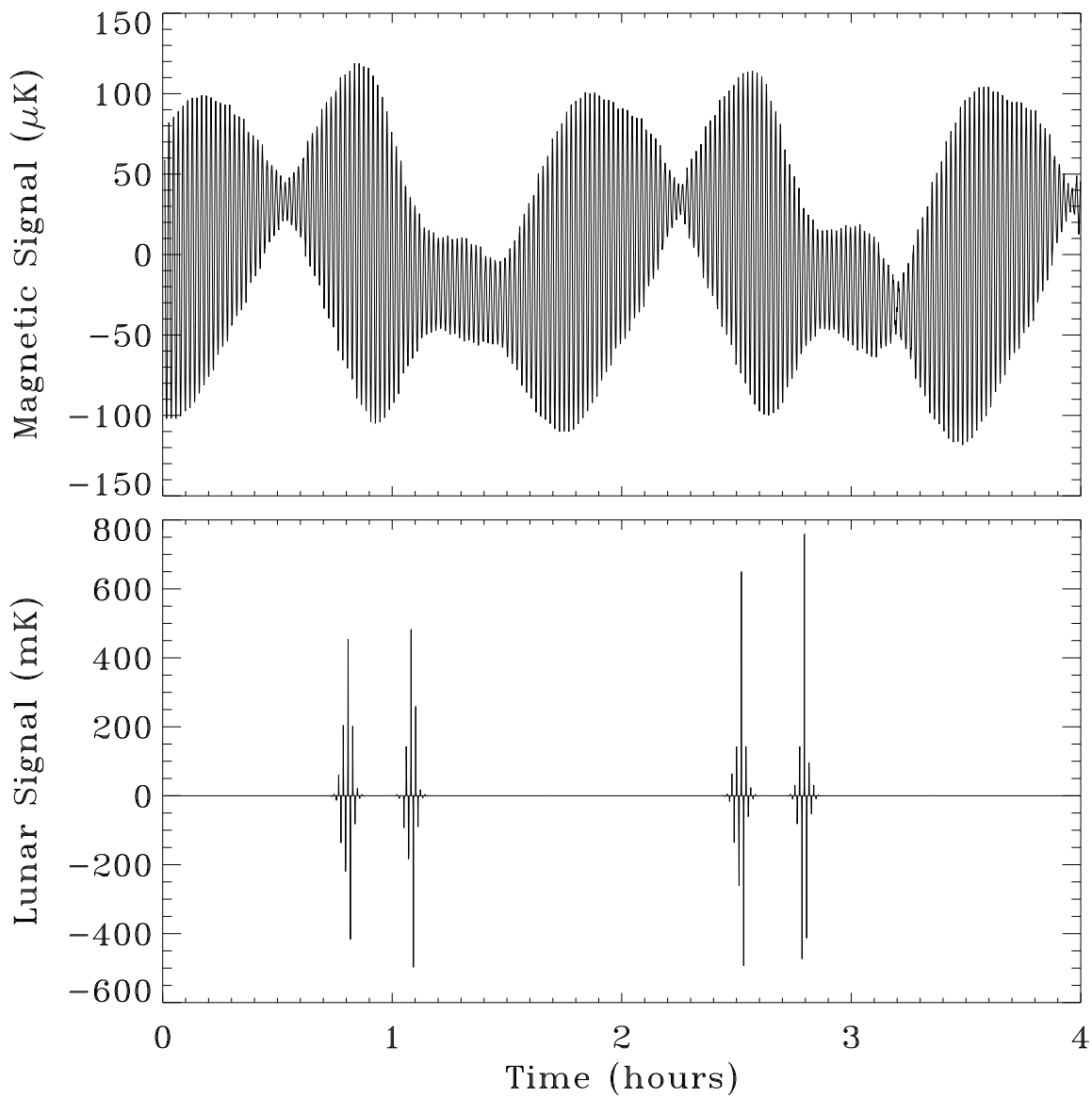}
\caption{
Magnetic and celestial signals vs time for 53B channel.
(top) Magnetic signal $Z_{\rm magnetic}(t)$ from the Earth's field.
The spin and orbit modulation are clearly apparent.
(bottom) Celestial signal from an unresolved source (the Moon).
The orbital modulation results from sweeping the antenna beam pattern
across the source at varying angles from beam center.
The magnetic signal is easily distinguished from fixed celestial sources.
}
\label{magsus_vs_time_fig}
\end{figure}

Figure \ref{magsus_vs_time_fig} shows the model magnetic signal
$Z_{\rm magnetic}(t)$ for for the 53B radiometer 
over the course of several orbits.
The spacecraft spin modulates both the radial and transverse field components,
causing the rapid variation at the spin period.
The spacecraft orbital motion samples the Earth's field at different latitudes, 
causing an orbital drift 
and modulating the amplitude envelope of the spin variations.
The resulting signal is quite distinct from that expected
from a fixed celestial signal,
represented in Figure \ref{magsus_vs_time_fig} by 
a model of lunar emission.
The different time-dependent signatures greatly reduce
the required accuracy of the magnetic model.

We simultaneously fit the time-ordered data in each channel
for the temperature in each pixel
and the magnetic susceptibility vector 
$\vec{\beta}$ 
(Eqs. \ref{chisq_eq} and \ref{magsus_eq}).
We obtain a significant improvement in $\chi^2$ in each channel,
demonstrating that 
the DMR magnetic signals are not well described by {\it any} 
set of fixed pixel temperatures.
More complicated models of the magnetic coupling
(e.g., tensor or non-linear terms)
do not further reduce the $\chi^2$:
a linear vector model (Eq.\ \ref{magsus_eq}) is sufficient.
We also fit 
$\vec{\beta}$ 
independently to each month of data
and search for time dependence in the fitted coefficients.
If the data are not corrected for the eclipse-related effects,
all channels show anomalies in the $\beta_X$ coefficients
during the eclipse season.
After correcting the time-ordered data for this effect ({\S}5.2),
the 31A and 31B channels
show additional modulation of the $\beta_X$ susceptibilities
with a period of one year.
We modify Eq.\ \ref{magsus_eq} to include additional terms
of the form 
$ \beta_{\rm annual} \cos\theta \cos\phi $
where $\theta$ is the orbit angle relative to the ascending node
and $\phi$ is the angle of the orbit plane relative to the vernal equinox.
Since the resulting signal is not spin-modulated,
it has almost no effect on the sky maps
but is included to reduce cross-talk with other orbitally modulated effects.
The 53 and 90 GHz channels show no significant variation 
in the fitted magnetic coefficients.

Table \ref{magsus_table} lists the magnetic coefficients
derived from 4 years of data.
Slow changes in the magnetic signal
from the orbitally-modulated $\beta_X$ susceptibility
may be removed as part of the baseline.
The values in Table \ref{magsus_table} 
refer to the two-orbit running mean baseline (Bennett et al.\ 1994)
for which no such subtraction takes place.

We model the magnetic field $\vec{B}(t)$
using the spacecraft attitude solution
and the time-dependent 
1985 International Geomagnetic Reference Field
(Barker et al.\ 1986)
to order $\ell = 8$.
We neglect the field components at higher $\ell$
and any contribution from the electromagnets used
to dissipate spacecraft angular momentum.
The amplitude of the field model
varies from 196 to 402 mG over the {\it COBE} orbit.
We limit deviations from the true field and our field model
by using on-board magnetometers 
and find good agreement
between the magnetometers and our application of the field model.
We have examined the residuals for coherent behavior with respect
to an inertial coordinate system
(e.g., if the electromagnets preferentially fired 
with the spacecraft in a fixed location and attitude)
and find none.
The {\it rms} difference between the magnetometers and the field model
is 8.9 mG averaged over the 4-year mission,
which we adopt as the 68\% CL uncertainty in the application of the model.

\begin{figure}[t]
\epsfxsize=6.0truein
\epsfbox{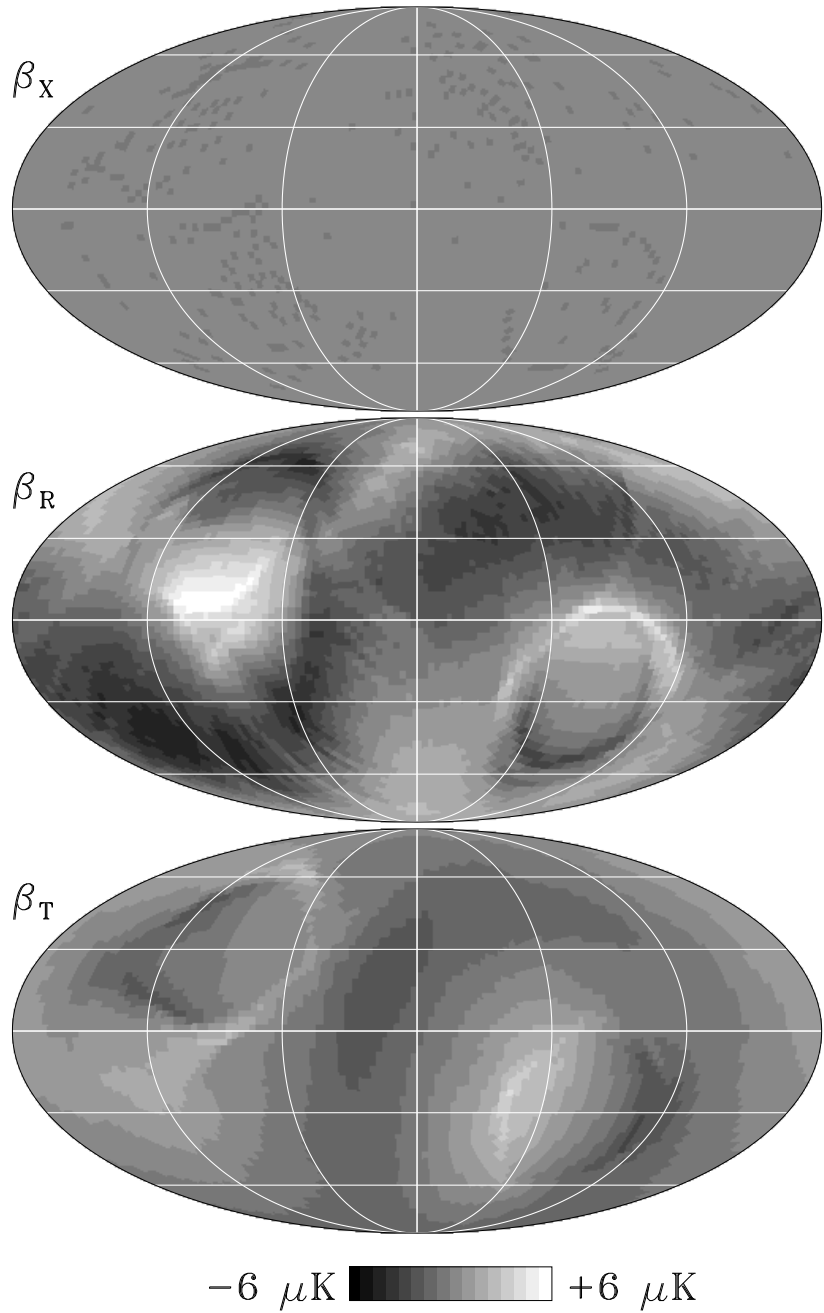}
\caption{
Full sky maps of 95\% CL upper limits to the residual effect,
after correction,
of the magnetic susceptibility in the 53B channel.
The maps are Mollweide projections in Galactic coordinates.
(top) $\beta_X$ susceptibility.
(middle) $\beta_R$ susceptibility.
(bottom) $\beta_T$ susceptibility.
A fitted dipole has been removed from each map to show
the higher-order structure.
}
\label{magsus_map_fig}
\end{figure}

The simultaneous fit for the pixel temperatures and magnetic coefficients
removes the fitted magnetic signals from the sky maps.
The uncertainties in the fitted coefficients are dominated
either by instrument noise
(if $\vec{\beta} \lsim 0.3 ~{\rm mK~G}^{-1}$)
or by the uncertainty in the magnetic field model.
We obtain a map of the removed signal
by subtracting the corrected sky map from a similar map 
for which no magnetic terms were fitted.
We estimate the residual uncertainties after correction
by multiplying this ``effect'' map
by the fractional uncertainty in each fitted coefficient
(Eq.\ \ref{syserr_map_eq}).
Figure \ref{magsus_map_fig} shows the 95\% CL uncertainties
in the 53B channel from magnetic effects.
The magnetic residuals are among the largest systematic uncertainties
in the DMR sky maps, but the amplitudes are small:
the residual magnetic uncertainty, after correction, 
is less than 3 \muK ~{\it rms} in any channel.

\subsection{Seasonal Effects}
Previous analysis of the 1- and 2-year data sets
showed the presence of an orbitally modulated signal
during the ``eclipse season'' surrounding the June solstice
when the spacecraft repeatedly flies through the Earth's shadow
(Kogut et al.\ 1992, Bennett et al.\ 1994).
This ``eclipse effect'' is strongly correlated
with various housekeeping signals,
particularly the 
unregulated spacecraft temperatures and bus voltages
which show the largest variation during eclipses and are thus
least affected by the telemetry digitization
(see Figure 4 of Kogut et al.\ 1992)
We model the effect empirically by fitting the time-ordered data to the form
$$
Z_{\rm eclipse} ~= ~a ~\Delta T_{\rm IPDU} ~+ ~b ~\Delta V_{\rm 28}
$$
where $\Delta T_{\rm IPDU}$ is the temperature of the IPDU box
and $\Delta V_{\rm 28}$ is the 28V bus voltage.
We remove an orbital mean from both 
$\Delta T_{\rm IPDU}$ and $\Delta V_{\rm 28U}$ prior to fitting
since long-term drifts are removed as part of the instrument baseline.
The resulting peak-peak changes in the housekeeping signals are
$\Delta T_{\rm IPDU} ~\sim$ 5.3 du,
$\Delta V_{\rm 28} ~\sim$ 4.3 du
during eclipse season, and
$\Delta T_{\rm IPDU} ~\sim$ 0.3 du,
$\Delta V_{\rm 28} ~\sim$ 0.1 du
excluding eclipse season,
where the housekeeping signals are expressed 
in digitized telemetry units (du).
Table \ref{eclipse_table} shows the fitted coefficients 
for this empirical model
during and excluding eclipse season.
We detect the effect in all channels during eclipse season.
Outside of eclipse season, 
the spacecraft temperatures and voltages are stable
and the effect vanishes.

The detection of thermal gain variations 
at the orbit period during eclipse season ({\S}4.2.2)
provides a plausible mechanism for the eclipse effect. 
However, neither the observed gain variations 
nor the empirical housekeeping correlations
remove the entire eclipse signal in all channels.
The magnetic $\beta_X$ coefficients for the 31A and 31B channels
remain anomalously large during eclipse season
even after correction for $Z_{\rm eclipse}$,
indicating that the empirical model removes only 2/3 of the signal 
in those channels.
The eclipse effect creates artifacts in the maps in two ways:
the direct projection of the signal into the maps,
and cross-talk with other orbitally modulated effects
(particularly the $\beta_X$ magnetic susceptibility).
Both effects are important for the 31A and 31B channels;
accordingly, we do not use data during eclipse season for these channels.
The 53 and 90 GHz channels have smaller eclipse coefficients
and show no $\beta_X$ anomalies after correction.
We correct the data using the empirical model
and estimate the residual uncertainties in the sky maps
using Eq. \ref{syserr_map_eq}.
Since the eclipse signal does not vary at the spin period,
its projection onto the maps is small.
Including the eclipse data in the 53 and 90 GHz channels
adds less than 0.3 \muK ~{\it rms} artifacts to the maps (95\% CL),
much less than the 10 \muK ~reduction in noise gained
by adding the 8 months of eclipse data to the 4-year data set.

\subsection{Orbit and Spin Effects}
\begin{figure}[t]
\epsfxsize=6.0truein
\epsfbox{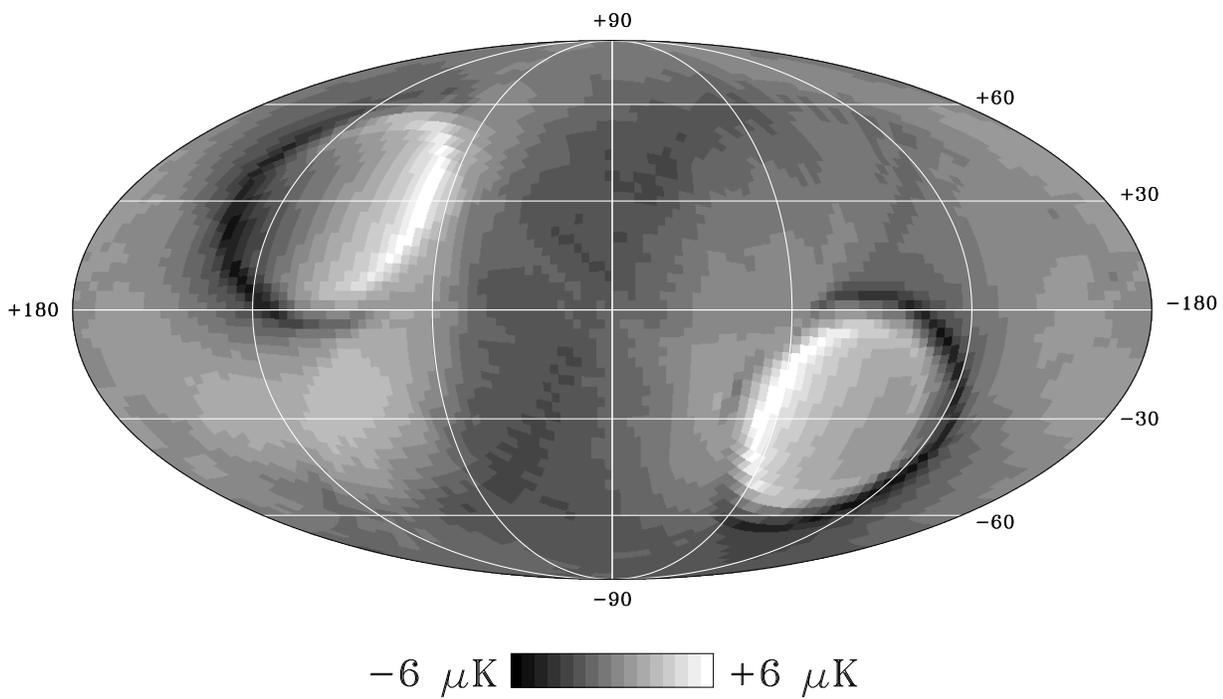}
\caption{
Full sky map of 95\% CL upper limits to 
spin-modulated effects in the 53B channel
(Mollweide projection in Galactic coordinates).
A fitted dipole has been removed to show the higher-order structure.
The amplitude of the structure in this map 
is dominated by the instrument noise binned at the spin period.
}
\label{spin_map_fig}
\end{figure}

The spacecraft orbit and spin provide a natural period for any environmental 
effects.  We test for the presence of additional effects,
independent of an {\it a priori} model,
by binning the corrected data $D_{ij}(t)$ by the
orbit angle with respect to the ascending node 
and the spin angle with respect to the solar vector.
With the exception of eclipse residuals
for the 31A and 31B channels,
there is no evidence for additional effects;
the binned data are compatible with instrument noise.
Table \ref{spin_orbit_table} shows the resulting upper limits
to the combined effects of any systematics at the spin and orbit periods,
after correction for magnetic and seasonal effects.
Artifacts in the maps from orbital effects at these levels are negligible
(below 0.3 \muK ~{\it rms} in any channel).
Synchronous effects at the spin period couple more strongly to the sky maps.
Figure \ref{spin_map_fig} shows the artifacts in the sky maps
resulting from spin-modulated signals
at the 95\% CL upper limit in Table \ref{spin_orbit_table}.
The noise limits on artifacts in the maps
resulting from combined effects at the spin period,
$\delta T_{\rm spin} < 1.6 ~\muK$ {\it rms},
are among the largest limits for the 4-year DMR maps.

\section{Foreground Sources}
Emission from foreground sources within the solar system
can create artifacts in maps of the microwave sky.
We reduce artifacts from foreground sources
by shielding the radiometers from the brightest sources (the Earth and Sun),
correcting the data using models of source microwave emission,
and rejecting data when uncertainties in the model
become unacceptably large (Table \ref{cut_list}).

\subsection{Earth}
The Earth is the largest foreground source,
emitting approximately 285 K over a quarter of the sky.
Emission from the Earth must be attenuated by 70 dB 
to reduce it below the 30 \muK ~level of typical CMB anisotropies.
DMR achieves this attenuation 
by using horn antennas with good off-axis sidelobe rejection
and by interposing a shield between the radiometers and the Earth.
Radiation from the Earth must diffract over the top of the shield
before affecting the DMR data.
The beam pattern, evaluated at the top of the shield,
is typically -65 dB or lower,
so only a modest attenuation from the shield is required.

We evaluate Earth emission by binning the time-ordered data $D_{ij}(t)$
by the position of the Earth limb
in a coordinate system fixed with respect to the {\it COBE} spacecraft.
Since the Earth subtends an angle 
much larger than the 7\deg ~DMR beam,
the azimuthal variation of the signal
with respect to the spacecraft spin
should reflect the antenna beam patterns,
with a positive lobe at the azimuth of horn 1 and a negative lobe
at the azimuth of horn 2 in each channel.
The signal change with respect to elevation angle
as the Earth sets below the shield
depends on the details of the diffraction over the shield edge,
while the overall normalization is set by the
beam response at the top of the shield.

For most of the mission, the attitude control
keeps the Earth well below the Sun/Earth shield.
During the eclipse season, the Earth rises as high 
as 8\deg ~above the shield.
When the Earth limb is above the shield, the binned data show a clear
detection of the expected dual-lobed signal.
At lower elevations, the signal decreases rapidly
and falls below the noise.
We derive limits to Earth emission at various elevation angles by
fitting the binned data to a model of Earth emission
based on scalar diffraction theory,
the relative geometry of the Earth, antennas, and shield,
and the measured beam patterns of each antenna.
Over narrow ranges of limb elevation angle
(e.g., a strip one pixel high),
the details of the diffraction become irrelevant
and only the azimuthal variation from the beam pattern is important.

\begin{figure}[t]
\epsfxsize=6.0truein
\epsfbox{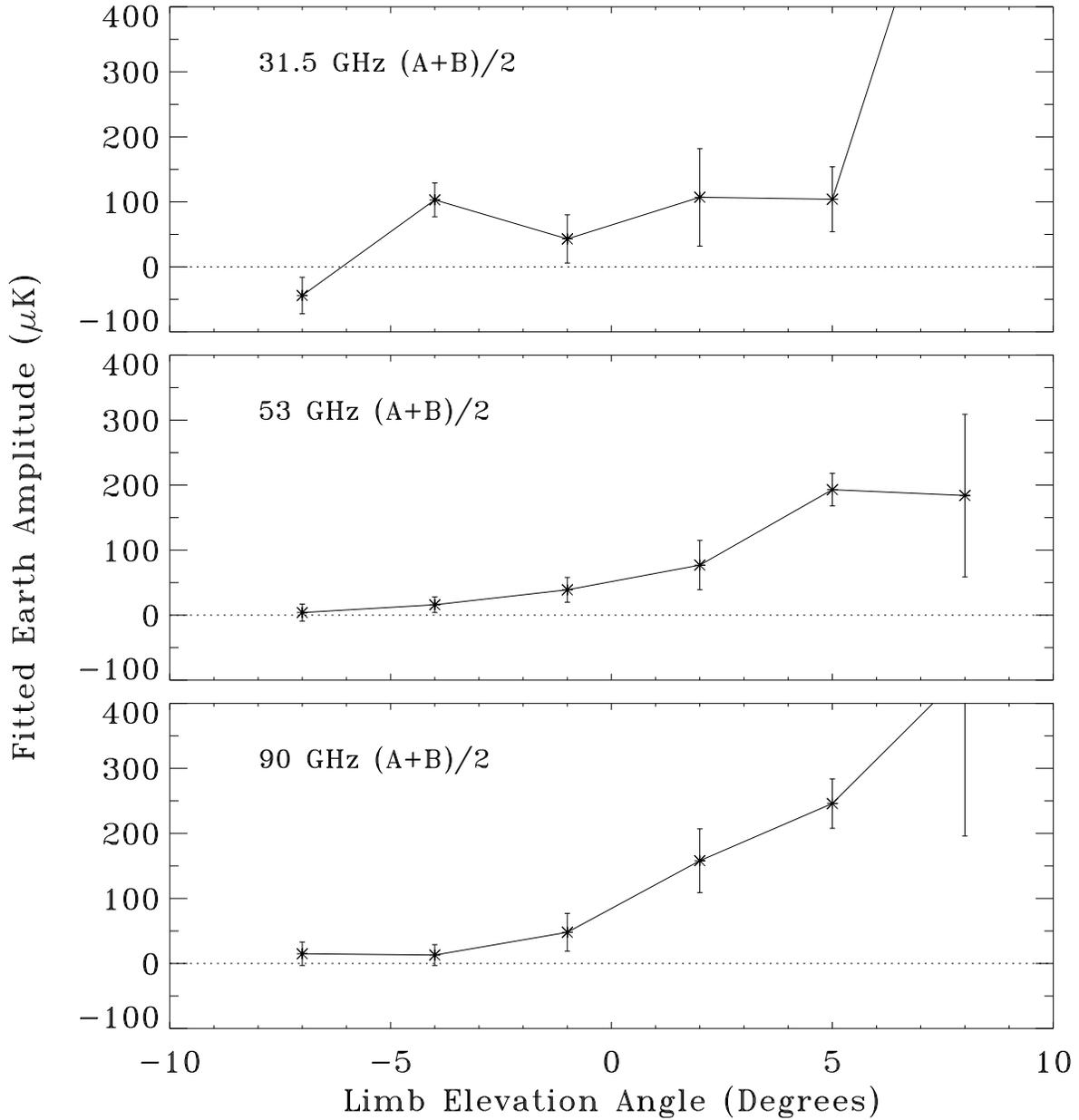}
\caption{
Amplitude of Earth emission in time-ordered data as a function
of Earth limb elevation angle.
The azimuthal variation from the beam patterns
has been fitted to the Earth-binned data
in a strip one pixel high.
Earth emission falls rapidly as the Earth sets below the shield.
We reject data for which the Earth limb is 1\deg ~below the shield or higher
(3\deg ~for the 31A and 31B channels).
}
\label{earth_vs_elev_fig}
\end{figure}

The model (with fitted amplitude as a free parameter)
provides a good fit to the binned data.
Figure \ref{earth_vs_elev_fig} 
shows the fitted amplitude versus elevation angle.
When the Earth is 5\deg ~above the shield,
we recover a fitted amplitude
$Z_{Earth} \approx 200 ~\muK$,
approximately 50\% of the expected amplitude.
This is well within the precision of the Earth model,
whose overall normalization depends on the exact position
of the deployed shield:
a 6 cm shift in shield position moves the shield edge 
as much as 4 dB in the antenna beam patterns.

We detect a mean signal $Z_{Earth} = 42 \pm 15 ~\muK$ when the
Earth is 1\deg ~below the shield,
falling to $1 \pm 10 ~\muK$ when the Earth is 7\deg ~below the shield.
This detection is somewhat larger than the upper limit
$Z_{Earth} < 30 ~\muK$ established from the 2-year data
(Bennett et al.\ 1994).
The difference results from Galactic emission in the 2-year data.
The Galactic plane and bright extended features
(Ophiuchus and Orion)
cross the antenna beam center
when the Earth is below the shield.
The resulting Galactic signal is brighter than the Earth emission,
even after binning in an Earth-based coordinate system.
Simulations using the DMR scan pattern
show that the binned Galactic emission
has the opposite phase as the Earth;
i.e., Galaxy crosses the positive antenna
when the Earth limb is at the azimuth of the negative antenna,
and partly cancels Earth emission
in the binned data.
We account for this effect in the 4-year binned data by
rejecting data with either antenna pointed at Galactic latitude $|b| < 15\deg$
and subtracting a Galactic model from the remaining high-latitude data.

\begin{figure}[t]
\epsfxsize=6.0truein
\epsfbox{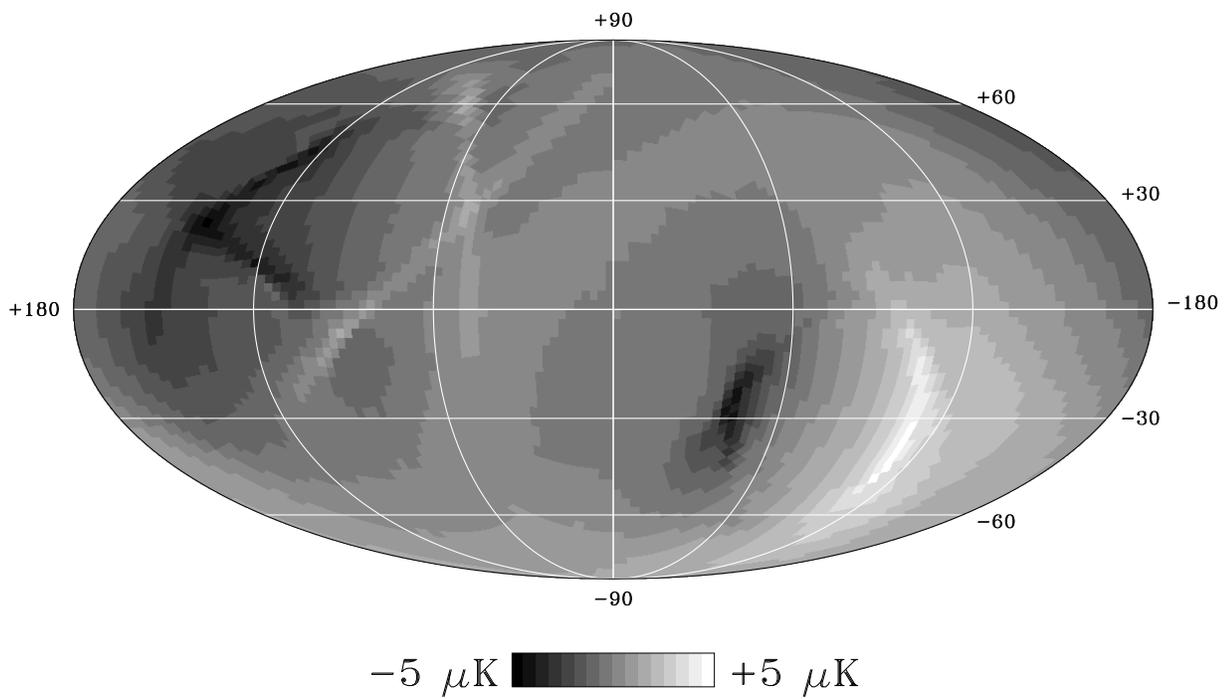}
\caption{
Full sky map of 95\% CL upper limits to 
Earth emission in the 53B channel
(Mollweide projection in Galactic coordinates).
}
\label{earth_map_fig}
\end{figure}

Table \ref{earth_table} shows the 95\% CL upper limits to Earth
emission when the Earth is 1\deg ~below the shield.
We reject data when the Earth limb is 1\deg ~below the shield or higher 
(3\deg ~for the 31.5 GHz channels)
but do not otherwise correct the data for the Earth.
The Earth signal decreases rapidly as the Earth sets below the shield
and falls below the threshold of detection 7\deg ~below the shield.
Earth emission is not detectable for the majority of the data set:
the Earth limb is 7\deg ~or more below the shield 
for 85\% of the 4-year mission.
We use the elevation dependence of the diffraction model 
to scale the upper limits at -1\deg ~to lower elevation angles,
and use this scaled model to map the Earth artifacts 
in the 4-year sky maps (Eqs. \ref{chisq_eq} and \ref{syserr_map_eq}).
Figure \ref{earth_map_fig} shows the 
95\% confidence level limits for Earth emission in the 53B sky map.
Since the upper limit is approximately twice the value of the detected signal,
the scaled limits at lower elevations are 
conservative estimates of Earth emission.
Earth emission contributes less than 1.8 \muK
~to the pixel-pixel {\it rms} in the sky maps.

An alternate approach is to use the
Earth-binned data as the model of Earth emission
without reference to any {\it a priori} model.
This avoids dependence on diffraction estimates
but injects the pixel noise of the binned data
into the mapping routine.
Limits set by this model-independent technique,
$\delta T_{Earth} < 3.5 ~\muK$,
are still small compared to the cosmic signal in the maps.

\subsection{Moon}
The Moon is the brightest source observed by DMR and is the
only source visible in the time-ordered data.
Away from beam center, 
its signal is rapidly attenuated by the antenna beam pattern.
The beam pattern falls to a local minimum of -39 dB 
at 21\deg ~from beam center,
at which point the Moon contributes approximately 150 \muK
~to the time-ordered data.
We reject any datum taken with an antenna pointed within 21\deg ~of the Moon
and correct all remaining data for lunar emission using an {\it a priori}
model based on lunar microwave emission properties and the measured
DMR beam patterns ({\S}4.1).  
Uncertainties in the lunar correction
are dominated by $\sim$5\% systematic uncertainties 
in the brightness temperature of the Moon,
and by $\sim$3\% uncertainties in the antenna beam pattern.
Residual uncertainties, after correction, are
less than 9 \muK ~in the time-ordered data
and less than 0.3 \muK ~in the maps.

\subsection{Sun}
A reflective shield protects the radiometers from the Sun,
which never directly illuminates the aperture plane.
A simple model of solar emission
diffracted over the shield
yields a limit
$Z_{\rm Sun} < 2 ~\muK$ in the time-ordered data (95\% CL).
We test for solar emission by binning the data by the solar position
relative to the spacecraft axes,
similar to the Earth binning in {\S}6.1.
We find no evidence for solar emission in the time-ordered data.
Solar effects are likely to be dominated 
by the heating of thermally-sensitive components 
and not by the direct microwave emission.
All solar effects (emission and thermal) are spin modulated
and are subsumed in the spin-binned limits in Table \ref{spin_orbit_table}
({\S} 5.3).

\subsection{Planets}
We correct the time-ordered data for emission from Jupiter, Saturn, and Mars
when those planets are within 15\deg ~of an antenna beam center.
Uncertainties in the applied corrections are dominated 
by $\sim$10\% uncertainties in the brightness temperature of each planet
(Kogut et al.\ 1992).
Residual artifacts in the maps, after correction,
are less than 0.4 \muK.
We test for planetary emission by mapping the time-ordered data
in a coordinate system centered on the planet Jupiter.
Accounting for the change in apparent diameter 
averaged over the 4 year mission,
Jupiter should appear in such a map as an unresolved source
with peak amplitude 200 \muK.
We fit the Jupiter-centered maps to a 7\deg ~FWHM Gaussian profile
and recover amplitudes 
$209 \pm 139 ~\mu$K, 
$195 \pm 40 ~\mu$K, and
$183 \pm 55 ~\mu$K
antenna temperature at 31.5, 53, and 90 GHz, respectively.

\subsection{RFI}
Radio-frequency interference (RFI) from ground-based radars,
if sufficiently strong,
will appear as spikes in the time-ordered data.
We have binned the flagged spikes by sub-satellite point and find
no correlation with position.
We test for RFI from geostationary satellites by
mapping the time-ordered data
in a coordinate system focused on the geostationary satellite belt,
including the effects of parallax.
Satellite RFI should appear as an unresolved source
on the equator of this coordinate system.
We find no such sources
to limit $Z_{\rm RFI} < 7 ~\muK$ in the time-ordered data
and $\delta T_{\rm RFI} < 0.02 ~\muK$ in the sky maps.

\section{Miscellaneous Effects}
Kogut et al.\ (1992) place stringent limits on a
large number of potential systematic artifacts
for the first year of DMR data.
We have repeated these analyses for the 4-year data
and include their effect in the combined limits
to all systematic artifacts;
however, we will not discuss each item separately.
These effects include
the solution convergence of the sparse matrix algorithm,
pixel-pixel independence,
discrete pixelization,
non-uniform sky coverage,
cross-talk with the DIRBE and FIRAS instruments,
emission from the Sun-Earth shield,
zodiacal dust emission,
and cosmic-ray hits.
Some of these effects are modulated at the spin or orbit periods
and are absorbed in the limits in Table \ref{spin_orbit_table}.

\subsection{Attitude Solution}
DMR uses the {\it COBE} fine aspect attitude solutions
based on {\it DIRBE} stellar observations.
For the first 45 months of the mission,
residuals between the attitude solution and known stellar positions
are less than 2\amin ~(68\% CL).
Fourier analysis of the residuals
shows no periodic systematic uncertainties
at either the spin or orbit periods;
the frequency spectrum is close to white noise.
A gyroscope failure on 2 October 1993
led to a wobble in spacecraft azimuth
of amplitude $\sim$13\amin
~which is not included in the attitude solutions.
We have simulated the effect of such a wobble in the 4-year DMR sky maps.
Attitude artifacts in the maps are less than 0.2 \muK 
~in the pixel-pixel {\it rms}.

\subsection{Antenna Direction Vectors}
The pointing of the 10 DMR antennas relative to the spacecraft body
was measured prior to launch.
We correct the pointing vectors of both 53B antennas
for a 0.25 s telemetry timing offset
(Kogut et al.\ 1992, Bennett et al.\ 1992a).
Observations of the Moon provide a cross-check on the
in-flight pointing of the antennas relative to the {\it COBE} spacecraft.
We find no offsets larger than 16\amin,
well within our ability to model the
centroid of the brightness distribution across the lunar disk.
We have made sky maps using the lunar-derived antenna vectors
and compared them to the mission maps made with the pre-flight
antenna vectors.
Systematic artifacts in the maps
related to antenna pointing vectors are less than 1.4 \muK.

\subsection{Lock-in Amplifier Memory}
We correct the time-ordered data for the small ``memory''
of the previous datum
caused by the low-pass filter on the lock-in amplifier input,
$$
D(t) = D(t_i) - \alpha D(t_{i-1})
$$
where the coefficient $\alpha \approx 0.032$
(Kogut et al.\ 1992).
In the absence of this correction,
the lock-in memory creates a positive correlation between
neighboring pixels in the sky maps.
We compute $\alpha$ for each channel
from the autocorrelation of the time-ordered data
after correcting the calibrated data for signals from the 
magnetic susceptibility,
eclipse effect,
lunar and planetary emission,
and the CMB and Earth Doppler dipoles.
Table \ref{lia_table} lists the resulting memory coefficients.
We simulate the effect of small uncertainties in the applied coefficients.
Artifacts in the maps resulting from uncertainties in the memory correction
are negligible (less than 0.05 \muK).

\section{Discussion}
We use Eqs. \ref{chisq_eq} and \ref{syserr_map_eq} to make a sky map 
of each effect before correction
and its 95\% confidence uncertainty after our best correction (if any)
is applied.
Note that the temperatures in these maps 
are highly correlated from pixel to pixel: 
the spatial pattern is fixed,
so that the residual (if any) across the DMR sky maps
will be given by scaling the entire upper limit map
in the range [-1, 1].

We expand each systematic error map in spherical harmonics,
$$
T(\theta, \phi) ~= ~\sum_{\ell m} ~a_{\ell m} Y_{\ell m}(\theta, \phi)
$$
and calculate the multipole amplitudes
$$
(\Delta T_\ell)^2 = \frac{1}{4 \pi} \sum_m | a_{\ell m} |^2
$$
for each map.
Tables \ref{31a_syserr_table} through \ref{90b_syserr_table}
present the {\it rms}, peak-peak amplitude,
and multipole amplitudes $\Delta T_\ell$ for each systematic effect.
The uncertainties in the potential systematic artifacts 
are largely independent;
we adopt the quadrature sum of each column as the 
upper limit for the combined effects of all systematics.
Note that this is not equivalent to adding the individual maps
and then deriving $\Delta T_\ell$.
Tables \ref{31a_syserr_table} through \ref{90b_syserr_table}
list only the largest uncertainties;
the row labelled ``Other'' contains the quadrature sum
for all other systematics.
This entry includes over a dozen individual effects,
such as 
attitude and pointing errors,
calibration drifts,
lock-in amplifier memory,
orbitally modulated effects,
planetary emission,
and artifacts from the map algorithm and pixelization.
See Kogut et al.\ (1992)
for a complete listing of these minor effects.

Upper limits on the combined effect of all systematics
in the pixel-pixel {\it rms} of the 4-year DMR sky maps
range from 5.4 \muK ~in the 31B channel
(which contains only 21 months of ``quiet'' data)
to 1.9 \muK ~in the 90B channel.
Upper limits on the {\it rms} quadrupole amplitude $\Delta T_2$
range from 3.8 \muK ~to 1.1 \muK.
The power drops rapidly at higher multipole moments $\ell$.

\begin{figure}[t]
\epsfxsize=6.0truein
\epsfbox{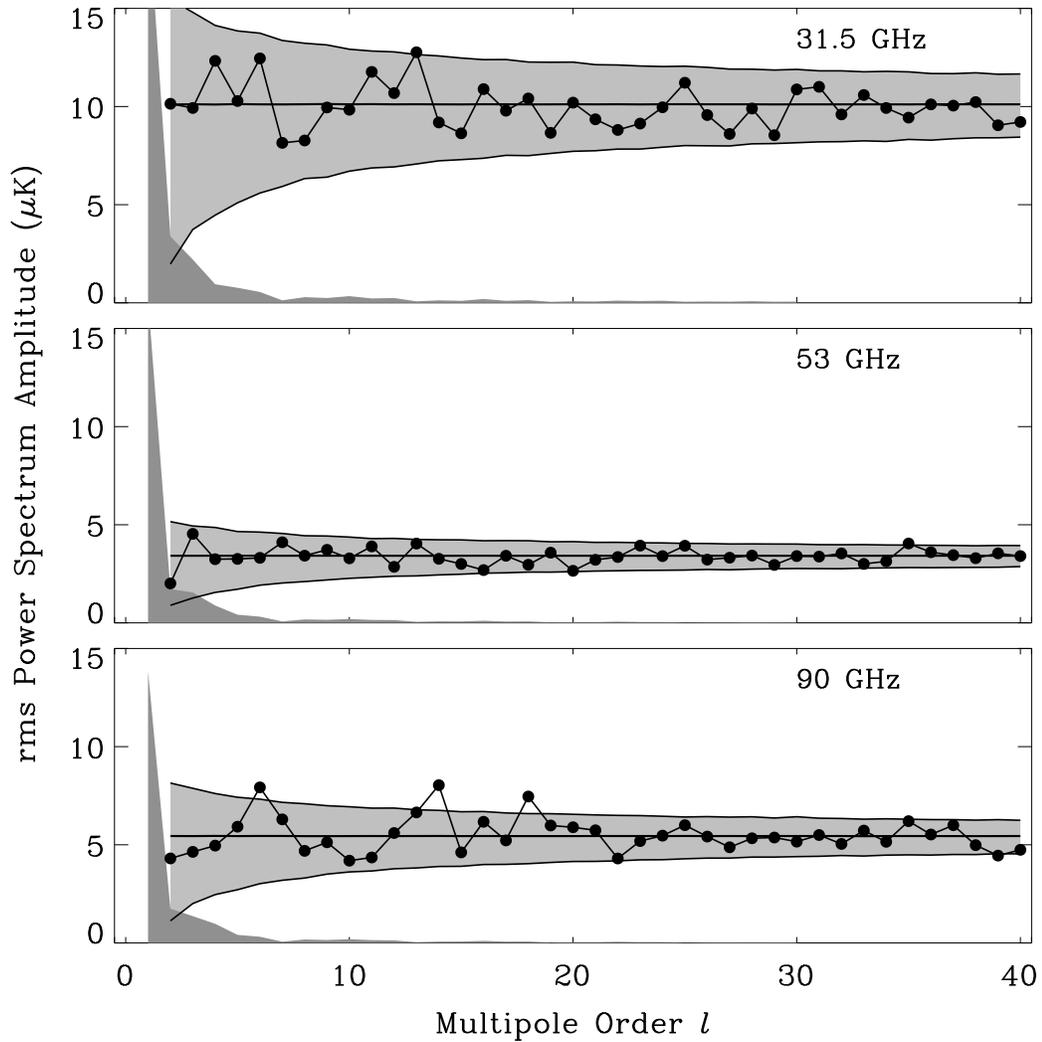}
\caption{
Power spectrum of the 4-year (A-B)/2 difference maps
compared to the instrument noise and upper limits on
the combined systematic uncertainties.
Points represent the power spectrum of the (A-B)/2 map,
while the light gray band represents the 95\% CL range 
of power from Monte Carlo simulations of instrument noise.
The dark gray band shows 95\% confidence level upper limits
to the quadrature sum of systematic effects.
}
\label{syserr_power_fig}
\end{figure}

We test for additional systematic artifacts in the sky maps
by analyzing various ``difference map'' linear combinations
in which celestial emission cancels,
leaving instrument noise and (potentially) systematic effects.
Examples are the (A-B)/2 differences between the
A and B channels at each frequency,
or similar maps made by differencing a single channel
mapped in different time ranges
(second year minus first year).
Figure \ref{syserr_power_fig} shows the power spectrum
of the 4-year (A-B)/2 difference maps
compared to the expected instrument noise and the
95\% confidence upper limits on combined systematic effects.
The gray band represents the 95\% CL range of power
from instrument noise,
determined by Monte Carlo simulation
using the observation pattern of each channel.
The power spectra of the difference maps
are in agreement with the expected instrument noise
and show no statistically significant excursions.
The upper limits to power from combined systematics
are well below the noise limits:
systematic artifacts do not limit analysis
of the DMR sky maps.

\section{Conclusions}
We use in-flight data from the full 4 year mission
to obtain estimates of potential systematic effects
in the DMR time-ordered data,
and use the mapping software to create maps of each effect
and its associated 95\% confidence level uncertainty.
The largest effect known to exist in the time-ordered data
is the instrument response to an external magnetic field.
We correct for this magnetic susceptibility using
a linear vector coupling between the radiometer orientation
and the Earth's magnetic field.
The uncertainty in the correction is dominated
by the instrument noise in the case of weak coupling
and uncertainty in the local magnetic field for larger couplings.

With four years of data, we detect emission from the Earth
at the level $42 \pm 15 ~\muK$ when the Earth is 1\deg
~below the Sun/Earth shield.
We reject data when the Earth is 1\deg ~below the shield or higher,
but do not otherwise correct for this emission.
Emission from the Earth agrees well with a 
simple diffraction model over the limited range of elevation angles
for which we detect the Earth.
We use the elevation dependence of the diffraction model
to scale the detection at -1\deg ~to lower elevation angles
when deriving upper limits to Earth artifacts in the 4-year sky maps.

We correct the calibrated time-ordered data for 
the known systematic effects (Table \ref{cut_list})
and bin the data by the spacecraft spin angle relative to the Sun.
We find no evidence for additional systematic effects
modulated at the spin period.
Our ability to detect weak signals with this method
is limited by the instrument noise;
upper limits to spin-modulated artifacts in the maps
determined from the spin-binned data are determined
by the instrument noise and not by a detected signal.

We detect an orbitally modulated signal in all channels
during the 2-month ``eclipse season''
surrounding the June solstice,
when the spacecraft flies through the Earth's shadow
over the Antarctic each orbit.
The detection of thermal gain variations
during the same time range
provides a plausible mechanism for this effect.
We correct the time-ordered data using an empirical model
based on variations in unregulated spacecraft temperature
and voltage signals,
which provide templates for the time variation
less affected by telemetry digitization
than the housekeeping signals on the amplifiers themselves.
The model removes about 2/3 of the signal.
The residual is large enough in the 31A and 31B channels
that the 4-year data would be adversely affected by including 
the corrected data with the residual signal.
We do not use data during the eclipse season for the 31A and 31B channels.

Observations of the Moon and the Doppler dipoles from the
motion of the satellite about the Earth and the Earth about the
Solar system barycenter
demonstrate that the pre-flight calibration is accurate
within its uncertainties.
On-board noise sources track the gain correctly
within 0.2\% per year.  Artifacts from calibration errors are negligible.
The 70 \muK ~Doppler signal from the orbital motion 
of the spacecraft about the Earth 
and the 200 \muK emission from Jupiter
serve as convenient test patterns
in the appropriate specialized coordinate systems.
DMR observes both signals with good signal to noise ratio.

We map the pattern of artifacts in the sky from the
95\% confidence level uncertainty in each potential systematic effect
and analyze these maps as though they were maps of the CMB.
The quadrature sum of the combined upper limits
ranges from 5.4 \muK ~to 1.9 \muK ~in the pixel-pixel {\it rms},
and from 3.8 \muK ~to 1.1 \muK ~for the {\it rms} quadrupole amplitude.
The power drops rapidly at higher multipole moments $\ell$.
The power spectra of the 4-year (A-B)/2 difference maps
are consistent with the distribution of instrument noise in the maps
and show no evidence for systematic artifacts.
Upper limits to power from combined systematics
in the DMR 4-year sky maps are well below the noise limits:
systematic artifacts do not limit analysis
of the DMR sky maps.

\vfill
\clearpage
\begin{center}
\large
{\bf References}
\end{center}

\normalsize
\pprintspace

\refitem
Barker, F.S., et al.\ 1986, EOS, 67, 523

\refitem
Bennett, C.L., et al.\ 1992a, ApJ, 391, 466

\refitem
--- 1992b, ApJ, 396, L7

\refitem
--- 1994, ApJ, 436, 423

\refitem
--- 1996, ApJ Letters, submitted

\refitem
Boggess, N.W., et al.\ 1992, ApJ, 397, 420

\refitem
Fixsen, D.J., Gales, J.M., Mather, J.C., Shafer, R.A., \& Wright, E.L. 1996,
ApJ, submitted

\refitem
Jackson, P.D., et al.\ 1996, in preparation

\refitem
--- 1992, in {\it Astronomical Data Analysis Software and Systems I},
ed. W. Worrall, C. Biemesderfer, and J. Barnes, (ASP:San Francisco), 382

\refitem
Janssen, M.A., \& Gulkis, S.\ 1992, in The Infrared and Submillimeter
Sky After {\it COBE}, ed.\ M.\ Signore \& C.\ Dupraz (Dordrecht: Kluwer), 391

\refitem
Lineweaver, C.H., Smoot, G.F., Bennett, C.L., Wright, E.L.,
Tenorio, L., Kogut, A., Keegstra, P.B., Hinshaw, G., \& Banday, A.J.\
1994, ApJ, 436, 452

\refitem
Keihm, S.J.\ 1982, Icarus, 53, 570

\refitem
Keihm, S.J.\ \& Gary, B.L.\ 1979, Proc.\ 10th Lun.\ Planet Sci.\ Conf., 2311

\refitem
Keihm, S.J.\ \& Langseth, M.G.\ 1975, Icarus, 24, 211

\refitem
Kogut, A., et al.\ 1992, ApJ, 401, 1

\refitem
---, Banday, A.J., Bennett, C.L., G\'{o}rski, K.M., Hinshaw, G., \& Reach, W.T.
1996a, ApJ, in press

\refitem
---, Hinshaw, G., Banday, A.J., Bennett, C.L., G\'{o}rski, K.M., \& Smoot, G.F.
1996b, APJ Letters, submitted


\refitem
Peebles, P.J.E., \& Wilkinson, D.T. 1968, Phys Rev, 174, 2168

\refitem
Toral, M.A., Ratliff, R.B., Lecha, M.C., Maruschak, J.G., Bennett, C.L., \& 
Smoot, G.F. 1990, IEEE Trans. Ant. Prop., 37, 171

\refitem
Torres, S., et al.\ 1989, in Data Analysis in Astronomy III,
ed.\ V.\ Di Gesu, L.\ Scarsi, \& M.C.\ Maccarone (New York: Plenum), 319

\refitem
Smoot, G.F., et al.\ 1990, ApJ, 360, 685

\refitem
--- et al.\ 1992, ApJ, 391, L1

\refitem
Wright, E.L., et al.\ 1994, ApJ, 420, 1

%
\vfill
\clearpage

\normalsize
\halfspace
\begin{table}
\caption{\label{cut_list}Cuts and Corrections Made to DMR Time-Ordered Data}
\begin{center}
\begin{tabular}{l l c}
\hline
Effect & Cut$^a$ & Correction \\
\hline
No telemetry			& Yes (0.4\%)		& No \\
Spike in data			& Yes (0.3\%)   	& No \\
Offscale data			& Yes (0.3\%)   	& No \\
Bad attitude			& Yes (0.7\%)   	& No \\
Unstable Noise			& Yes (4.2\%)$^b$	& No \\
Calibration			& Yes (2.2\%)   	& No \\
Earth emission			& Yes (5.0\%)   	& No \\
Lunar emission			& Yes (4.6\%)   	& Yes \\
Planetary emission		& No			& Yes \\
Lock-in amplifier memory	& No			& Yes \\
Magnetic susceptibility		& No			& Yes \\
Earth Doppler			& No			& Yes \\
Satellite Doppler		& No			& Yes \\
Seasonal effects		& Yes (26.0\%)$^c$	& Yes \\
\hline
\end{tabular}
\end{center}
$^a$ Numbers in parentheses are the fraction of data rejected for each cut. \\
$^b$ 31B channel only. \\
$^c$ 31A and 31B channels only.
\end{table}

\normalsize
\halfspace
\begin{table}
\caption{\label{cal_summary}Calibration Summary$^a$}
\begin{center}
\begin{tabular}{l c r c r c c}
\hline
Channel & \multicolumn{2}{c}{Absolute Calibration} & Relative & 
\multicolumn{1}{c}{Linear Drift} & Orbit Drift & Spin Drift \\
 & Ground (\%) & \multicolumn{1}{c}{Flight (\%)} & A/B (\%) & 
\multicolumn{1}{c}{(percent yr$^{-1}$)} & 
$( 10^5 ~\Delta {\cal G}/{\cal G} )$ &
$( 10^7 ~\Delta {\cal G}/{\cal G} )$ \\
\hline
31A & 0.0 $\pm$ 2.5 & +2.2 $\pm$ 3.0 & 0.40 $\pm$ 0.05
	& +0.23 $\pm$ 0.33 &  6.7 & 15 \\
31B & 0.0 $\pm$ 2.3 & +1.6 $\pm$ 3.8 & 
	& +0.03 $\pm$ 0.81 & 22.5 & 55 \\
53A & 0.0 $\pm$ 0.7 & -0.8 $\pm$ 1.2 & 0.27 $\pm$ 0.02
	& -0.12 $\pm$ 0.03 &  2.6 & 20 \\
53B & 0.0 $\pm$ 0.7 & -0.1 $\pm$ 1.4 & 
	& -0.11 $\pm$ 0.11 &  2.4 & 13 \\
90A & 0.0 $\pm$ 2.0 & +2.6 $\pm$ 2.5 & 0.01 $\pm$ 0.03
	& -0.12 $\pm$ 0.04 &  5.4 & 13 \\
90B & 0.0 $\pm$ 1.3 & -1.7 $\pm$ 1.8 & 
	& -0.20 $\pm$ 0.11 &  7.2 &  8 \\
\hline
\end{tabular}
\end{center}
$^a$ These corrections to the ground calibration have {\it not}
been applied to the data.  Uncertainties are 68\% confidence level
except for the spin and orbit drifts, which are 95\% confidence upper limits.
\end{table}

\normalsize
\halfspace
\begin{table}
\caption{\label{abs_cal}Absolute Calibration Compared to Noise Sources}
\begin{center}
\begin{tabular}{l c c c}
\hline
Channel & Ground & Doppler & Moon \\
\hline
31A & 1.000 $\pm$ 0.025 & 1.022 $\pm$ 0.030 & 1.021 $\pm$ 0.038 \\
31B & 1.000 $\pm$ 0.023 & 1.116 $\pm$ 0.058 & 1.016 $\pm$ 0.038 \\
53A & 1.000 $\pm$ 0.007 & 0.992 $\pm$ 0.012 & 1.021 $\pm$ 0.054 \\
53B & 1.000 $\pm$ 0.007 & 0.999 $\pm$ 0.014 & 1.018 $\pm$ 0.054 \\
90A & 1.000 $\pm$ 0.020 & 1.026 $\pm$ 0.025 & 1.014 $\pm$ 0.063 \\
90B & 1.000 $\pm$ 0.013 & 0.983 $\pm$ 0.018 & 1.013 $\pm$ 0.063 \\
\hline
\end{tabular}
\end{center}
\end{table}

\normalsize
\halfspace
\begin{table}
\caption{\label{cross_cal}Relative A/B Calibration Compared to Noise Sources}
\begin{center}
\begin{tabular}{l c c c}
\hline
Frequency & Ground & Doppler & Moon \\
\hline
31 & 1.000 $\pm$ 0.026 & 0.915 $\pm$ 0.054 & 1.0040 $\pm$ 0.0005 \\
53 & 1.000 $\pm$ 0.003 & 0.993 $\pm$ 0.018 & 1.0027 $\pm$ 0.0002 \\
90 & 1.000 $\pm$ 0.010 & 1.044 $\pm$ 0.032 & 1.0001 $\pm$ 0.0003 \\
\hline
\end{tabular}
\end{center}
\end{table}

\normalsize
\halfspace
\begin{table}
\caption{\label{linear_drifts}Linear Calibration Drifts}
\begin{center}
\begin{tabular}{l c c c}
\hline
Channel & NS Ratio & CMB Dipole & Moon \\
 & (percent yr$^{-1}$)  & (percent yr$^{-1}$)  & (percent yr$^{-1}$) \\
\hline
31A & +0.468 $\pm$ 0.004 & +0.365 $\pm$ 0.278 & -0.140 $\pm$ 0.024 \\
31B & -0.576 $\pm$ 0.015 & +0.947 $\pm$ 1.281 & -0.293 $\pm$ 0.029 \\
53A & -0.139 $\pm$ 0.001 & -0.091 $\pm$ 0.122 & -0.131 $\pm$ 0.009 \\
53B & -0.191 $\pm$ 0.001 & +0.023 $\pm$ 0.139 & -0.147 $\pm$ 0.011 \\
90A & -0.099 $\pm$ 0.003 & -0.173 $\pm$ 0.209 & -0.098 $\pm$ 0.018 \\
90B & -0.131 $\pm$ 0.001 & -0.325 $\pm$ 0.174 & -0.140 $\pm$ 0.015 \\
\hline
\end{tabular}
\end{center}
\end{table}

\normalsize
\halfspace
\begin{table}
\caption{\label{magsus_table}Magnetic Susceptibility (mK G$^{-1}$)}
\begin{center}
\begin{tabular}{l r r r}
\hline
Channel & 
\multicolumn{1}{c}{$\beta_X$} & 
\multicolumn{1}{c}{$\beta_R$} & 
\multicolumn{1}{c}{$\beta_T$} \\
\hline
31A & -0.173 $\pm$ 0.018 &  +0.262 $\pm$ 0.039 &  -0.196 $\pm$ 0.067 \\
31B & +0.284 $\pm$ 0.024 &  +0.224 $\pm$ 0.054 &  +0.011 $\pm$ 0.089 \\
53A & -1.514 $\pm$ 0.006 &  -0.081 $\pm$ 0.013 &  -0.881 $\pm$ 0.022 \\
53B & +0.087 $\pm$ 0.006 &  -0.408 $\pm$ 0.015 &  -0.192 $\pm$ 0.025 \\
90A & -0.135 $\pm$ 0.010 &  -1.175 $\pm$ 0.024 &  -0.334 $\pm$ 0.040 \\
90B & -0.003 $\pm$ 0.007 &  +0.140 $\pm$ 0.017 &  -0.136 $\pm$ 0.028 \\
\hline
\end{tabular}
\end{center}
\end{table}

\normalsize
\halfspace
\begin{table}
\caption{\label{eclipse_table}Eclipse Coefficients for Empirical Model}
\begin{center}
\begin{tabular}{l r r r r}
\hline
Channel & \multicolumn{2}{c}{Thermal ( mK du$^{-1}$)$^a$} & 
\multicolumn{2}{c}{Voltage (mK du$^{-1}$)$^a$} \\
 & 
\multicolumn{1}{c}{During Eclipse} & 
\multicolumn{1}{c}{Excluding Eclipse} & 
\multicolumn{1}{c}{During Eclipse} & 
\multicolumn{1}{c}{Excluding Eclipse} \\
\hline
31A & +0.442 $\pm$ 0.006 & +0.046 $\pm$ 0.029 &
      +0.311 $\pm$ 0.015 & +0.018 $\pm$ 0.019 \\
31B & +0.147 $\pm$ 0.008 & -0.011 $\pm$ 0.038 &
      +0.262 $\pm$ 0.019 & -0.008 $\pm$ 0.027 \\
53A & -0.022 $\pm$ 0.002 & +0.001 $\pm$ 0.007 &
      +0.013 $\pm$ 0.004 & -0.002 $\pm$ 0.007 \\
53B & -0.069 $\pm$ 0.003 & -0.010 $\pm$ 0.010 &
      -0.011 $\pm$ 0.004 & +0.001 $\pm$ 0.007 \\
90A & +0.042 $\pm$ 0.004 & +0.032 $\pm$ 0.018 &
      +0.024 $\pm$ 0.008 & +0.009 $\pm$ 0.011 \\
90B &  0.000 $\pm$ 0.003 & -0.018 $\pm$ 0.011 &
       0.000 $\pm$ 0.005 & +0.008 $\pm$ 0.011 \\
\hline
\end{tabular}
\end{center}
$^a$~Temperature and voltage housekeeping signals are
processed in digitized telemetry units (du).
\end{table}

\normalsize
\halfspace
\begin{table}
\caption{\label{spin_orbit_table}
95\% Confidence Upper Limits 
From Orbit- and Spin-Modulated Effects$^a$}
\begin{center}
\begin{tabular}{l c c c}
\hline
Channel & Spin Period & Orbit Period   & Orbit Period \\
        & All Data    & During Eclipse & Excluding Eclipse \\
        & (\muK)      & (\muK)         & (\muK) \\
\hline
31A & 	34	&  577	&  19 \\
31B & 	41	&  489	&  47 \\
53A & 	11	&   33	&  14 \\
53B & 	10	&   19	&  12 \\
90A & 	18	&   73 	&  24 \\
90B & 	8	&   21	&  18 \\
\hline
\end{tabular}
\end{center}
$^a$~All values are in units of antenna temperature.
\end{table}

\normalsize
\halfspace
\begin{table}
\caption{\label{earth_table}
95\% Confidence Upper Limits 
From Earth Emission in Time-Ordered Data$^a$}
\begin{center}
\begin{tabular}{l c c}
\hline
Channel & 1\deg ~below shield$^b$ & 7\deg ~below shield$^c$ \\
        & (\muK) & (\muK) \\
\hline
31A & 179 & 26 \\
31B & 183 & 27 \\
53A &  85 & 8 \\
53B &  89 & 10 \\
90A & 136 & 10 \\
90B & 113 & 7 \\
\hline
\end{tabular}
\end{center}
$^a$~All values are in units of antenna temperature. \\
$^b$~Direct fit to binned data 1\deg ~below the shield. \\
$^c$~Values at -7\deg ~are taken from the values at -1\deg ~(column (2)),
scaled to -7\deg ~using diffraction model. \\
\end{table}

\normalsize
\halfspace
\begin{table}
\caption{\label{lia_table}
Lock-in Amplifier Memory}
\begin{center}
\begin{tabular}{l c}
\hline
Channel & Lock-in Memory Amplitude \\
        & (percent of signal) \\
\hline
31A & 3.220 $\pm$ 0.003 \\
31B & 3.146 $\pm$ 0.004 \\
53A & 3.203 $\pm$ 0.003 \\
53B & 3.172 $\pm$ 0.003 \\
90A & 3.110 $\pm$ 0.003 \\
90B & 3.139 $\pm$ 0.004 \\
\hline
\end{tabular}
\end{center}
\end{table}

\normalsize
\halfspace
\begin{table}
\caption{\label{31a_syserr_table}
Systematic Effects for Channel 31A$^a$}
\begin{center}
\begin{tabular}{l r r r r r r r r r r}
\hline
Effect & P-P$^b$ & rms$^c$ & $\Delta T_1$ & $\Delta T_2$ & $\Delta T_3$ 
       & $\Delta T_4$ & $\Delta T_5$ & $\Delta T_6$ & $\Delta T_7$ 
       & $\Delta T_8$ \\
       & ($\mu$K) & ($\mu$K) & ($\mu$K) & ($\mu$K) & ($\mu$K) & ($\mu$K) 
       & ($\mu$K) & ($\mu$K) & ($\mu$K) & ($\mu$K) \\
\hline
\multicolumn{11}{c}{Channel 31A Before Correction} \\
\hline
$\beta_X$  &   6.7   &   0.4   &   0.2   &   0.1   &   0.1   &   0.0  
    &   0.1   &   0.0   &   0.0   &   0.0   \\
$\beta_R$  &  31.9   &   6.5   &   1.1   &   6.1   &   0.5   &   1.6  
    &   0.7   &   1.0   &   0.2   &   0.5   \\
$\beta_T$  &  17.4   &   2.9   &  18.4   &   1.6   &   1.9   &   0.7  
    &   0.9   &   0.3   &   0.1   &   0.2   \\
Earth      &   7.0   &   1.2   &   1.5   &   0.9   &   0.4   &   0.4  
    &   0.3   &   0.2   &   0.0   &   0.1   \\
Moon       &   7.9   &   1.6   &   0.0   &   1.2   &   0.0   &   0.4  
    &   0.1   &   0.3   &   0.0   &   0.5   \\
Doppler    &  85.6   &  13.4   &  65.3   &   5.6   &  10.4   &   1.8  
    &   3.5   &   1.4   &   0.8   &   0.7   \\
Spin       &  11.1   &   1.5   &   5.3   &   0.6   &   1.0   &   0.4  
    &   0.4   &   0.3   &   0.0   &   0.2   \\
Other      &  81.3   &   9.1   &   2.3   &   0.8   &   0.4   &   0.5  
    &   0.4   &   0.5   &   0.4   &   0.5   \\
 & & & & & & & & & & \\
Total$^d$  & 124.7   &  17.9   &  68.2   &   8.6   &  10.6   &   2.7  
    &   3.7   &   1.8   &   0.9   &   1.2   \\
\hline
\multicolumn{11}{c}{Channel 31A After Correction$^e$} \\
\hline
$\beta_X$  &   1.6   &   0.1   &   0.1   &   0.0   &   0.0   &   0.0  
    &   0.0   &   0.0   &   0.0   &   0.0   \\
$\beta_R$  &  10.3   &   2.1   &   0.4   &   2.0   &   0.2   &   0.5  
    &   0.2   &   0.3   &   0.1   &   0.2   \\
$\beta_T$  &  12.1   &   2.0   &  12.8   &   1.1   &   1.3   &   0.5  
    &   0.6   &   0.2   &   0.1   &   0.1   \\
Earth      &   7.6   &   1.3   &   1.6   &   1.0   &   0.4   &   0.4  
    &   0.3   &   0.2   &   0.0   &   0.2   \\
Moon       &   0.9   &   0.2   &   0.0   &   0.1   &   0.0   &   0.1  
    &   0.0   &   0.0   &   0.0   &   0.1   \\
Doppler    &   4.3   &   0.7   &   3.3   &   0.3   &   0.5   &   0.1  
    &   0.2   &   0.1   &   0.0   &   0.0   \\
Spin       &  23.2   &   3.1   &  11.1   &   1.3   &   2.0   &   0.8  
    &   0.8   &   0.7   &   0.1   &   0.4   \\
Other      &  15.6   &   1.4   &   2.5   &   0.5   &   0.5   &   0.3  
    &   0.2   &   0.2   &   0.1   &   0.3   \\
 & & & & & & & & & & \\
Total$^d$  &  33.4   &   4.7   &  17.5   &   2.8   &   2.5   &   1.2  
    &   1.1   &   0.9   &   0.2   &   0.6   \\
\hline
\end{tabular}
\end{center}
$^a$~All results are in units of antenna temperature.\\
$^b$~Peak to peak amplitude in the map after best-fit dipole is removed.\\
$^c$~Pixel to pixel standard deviation after best-fit dipole is removed.\\
$^d$~Quadrature sum of the individual effects in each column.\\
$^e$~95\% confidence upper limits.\\
\end{table}

\normalsize
\halfspace
\begin{table}
\caption{\label{31b_syserr_table}
Systematic Effects for Channel 31B$^a$}
\begin{center}
\begin{tabular}{l r r r r r r r r r r}
\hline
Effect & P-P & rms & $\Delta T_1$ & $\Delta T_2$ & $\Delta T_3$ 
       & $\Delta T_4$ & $\Delta T_5$ & $\Delta T_6$ & $\Delta T_7$ 
       & $\Delta T_8$ \\
       & ($\mu$K) & ($\mu$K) & ($\mu$K) & ($\mu$K) & ($\mu$K) & ($\mu$K) 
       & ($\mu$K) & ($\mu$K) & ($\mu$K) & ($\mu$K) \\
\hline
\multicolumn{11}{c}{Channel 31B Before Correction} \\
\hline
$\beta_X$  &   8.4   &   0.5   &   0.3   &   0.1   &   0.1   &   0.1  
    &   0.1   &   0.1   &   0.0   &   0.0   \\
$\beta_R$  &  26.3   &   5.1   &   0.6   &   4.5   &   0.7   &   1.6  
    &   0.6   &   0.5   &   0.1   &   0.4   \\
$\beta_T$  &   0.0   &   0.0   &   0.0   &   0.0   &   0.0   &   0.0  
    &   0.0   &   0.0   &   0.0   &   0.0   \\
Earth      &  11.5   &   1.6   &   0.5   &   1.5   &   0.3   &   0.3  
    &   0.3   &   0.2   &   0.0   &   0.1   \\
Moon       &   8.5   &   1.6   &   0.1   &   1.3   &   0.0   &   0.4  
    &   0.0   &   0.3   &   0.0   &   0.5   \\
Doppler    &  97.1   &  15.6   &  66.5   &   6.1   &  12.0   &   2.7  
    &   4.6   &   2.2   &   1.2   &   1.2   \\
Spin       &   8.6   &   1.1   &   4.1   &   0.5   &   0.8   &   0.3  
    &   0.3   &   0.3   &   0.0   &   0.2   \\
Other      &  95.9   &  10.9   &   0.7   &   0.7   &   0.3   &   0.5  
    &   0.3   &   0.5   &   0.4   &   0.5   \\
 & & & & & & & & & & \\
Total      & 140.2   &  19.9   &  66.6   &   7.9   &  12.1   &   3.2  
    &   4.7   &   2.3   &   1.3   &   1.5   \\
\hline
\multicolumn{11}{c}{Channel 31B After Correction} \\
\hline
$\beta_X$  &   1.7   &   0.1   &   0.1   &   0.0   &   0.0   &   0.0  
    &   0.0   &   0.0   &   0.0   &   0.0   \\
$\beta_R$  &  13.1   &   2.5   &   0.3   &   2.3   &   0.4   &   0.8  
    &   0.3   &   0.3   &   0.1   &   0.2   \\
$\beta_T$  &   0.0   &   0.0   &   0.0   &   0.0   &   0.0   &   0.0  
    &   0.0   &   0.0   &   0.0   &   0.0   \\
Earth      &  13.0   &   1.8   &   0.6   &   1.6   &   0.4   &   0.3  
    &   0.3   &   0.2   &   0.1   &   0.1   \\
Moon       &   0.9   &   0.2   &   0.0   &   0.1   &   0.0   &   0.1  
    &   0.0   &   0.0   &   0.0   &   0.1   \\
Doppler    &   4.5   &   0.7   &   3.1   &   0.3   &   0.6   &   0.1  
    &   0.2   &   0.1   &   0.1   &   0.1   \\
Spin       &  28.2   &   3.7   &  13.4   &   1.5   &   2.5   &   0.9  
    &   1.0   &   0.9   &   0.1   &   0.5   \\
Other      &  22.1   &   2.6   &   2.8   &   0.5   &   1.1   &   0.4  
    &   0.5   &   0.3   &   0.2   &   0.3   \\
 & & & & & & & & & & \\
Total      &  40.6   &   5.6   &  14.1   &   3.2   &   2.8   &   1.4  
    &   1.2   &   1.0   &   0.2   &   0.6   \\
\hline
\end{tabular}
\end{center}
$^a$~All results are in units of antenna temperature.
Columns are the same as Table \ref{31a_syserr_table}.\\
\end{table}

\normalsize
\halfspace
\begin{table}
\caption{\label{53a_syserr_table}
Systematic Effects for Channel 53A$^a$}
\begin{center}
\begin{tabular}{l r r r r r r r r r r}
\hline
Effect & P-P & rms & $\Delta T_1$ & $\Delta T_2$ & $\Delta T_3$ 
       & $\Delta T_4$ & $\Delta T_5$ & $\Delta T_6$ & $\Delta T_7$ 
       & $\Delta T_8$ \\
       & ($\mu$K) & ($\mu$K) & ($\mu$K) & ($\mu$K) & ($\mu$K) & ($\mu$K) 
       & ($\mu$K) & ($\mu$K) & ($\mu$K) & ($\mu$K) \\
\hline
\multicolumn{11}{c}{Channel 53A Before Correction} \\
\hline
$\beta_X$  &  32.9   &   2.2   &   5.1   &   0.4   &   0.8   &   0.4  
    &   0.4   &   0.3   &   0.1   &   0.1   \\
$\beta_R$  &   6.3   &   1.0   &   0.6   &   0.6   &   0.2   &   0.7  
    &   0.2   &   0.2   &   0.0   &   0.1   \\
$\beta_T$  & 101.0   &  15.7   &  92.8   &   7.6   &  11.9   &   3.0  
    &   2.7   &   1.7   &   0.5   &   1.0   \\
Earth      &   6.6   &   0.7   &   0.5   &   0.6   &   0.2   &   0.2  
    &   0.1   &   0.1   &   0.0   &   0.1   \\
Moon       &   7.1   &   1.5   &   0.0   &   1.2   &   0.0   &   0.4  
    &   0.0   &   0.3   &   0.0   &   0.5   \\
Doppler    &  70.9   &  10.9   &  50.5   &   6.8   &   6.7   &   1.8  
    &   3.0   &   1.9   &   0.5   &   1.3   \\
Spin       &   3.3   &   0.4   &   1.6   &   0.2   &   0.3   &   0.1  
    &   0.1   &   0.1   &   0.0   &   0.1   \\
Other      &  37.3   &   3.7   &   1.0   &   1.0   &   0.4   &   0.6  
    &   0.3   &   0.5   &   0.4   &   0.5   \\
 & & & & & & & & & & \\
Total      & 133.6   &  19.7   & 105.7   &  10.4   &  13.7   &   3.7  
    &   4.1   &   2.6   &   0.8   &   1.8   \\
\hline
\multicolumn{11}{c}{Channel 53A After Correction} \\
\hline
$\beta_X$  &   4.3   &   0.3   &   0.7   &   0.1   &   0.1   &   0.1  
    &   0.1   &   0.0   &   0.0   &   0.0   \\
$\beta_R$  &   2.2   &   0.4   &   0.2   &   0.2   &   0.1   &   0.3  
    &   0.1   &   0.1   &   0.0   &   0.0   \\
$\beta_T$  &  14.3   &   2.2   &  13.2   &   1.1   &   1.7   &   0.4  
    &   0.4   &   0.2   &   0.1   &   0.2   \\
Earth      &   7.9   &   0.9   &   0.6   &   0.8   &   0.2   &   0.2  
    &   0.1   &   0.1   &   0.0   &   0.1   \\
Moon       &   0.9   &   0.2   &   0.0   &   0.1   &   0.0   &   0.1  
    &   0.0   &   0.0   &   0.0   &   0.1   \\
Doppler    &   1.2   &   0.2   &   0.7   &   0.1   &   0.1   &   0.0  
    &   0.1   &   0.0   &   0.0   &   0.0   \\
Spin       &   7.6   &   1.0   &   3.6   &   0.4   &   0.7   &   0.3  
    &   0.3   &   0.2   &   0.0   &   0.1   \\
Other      &  10.6   &   1.0   &   0.9   &   0.7   &   0.1   &   0.2  
    &   0.1   &   0.2   &   0.1   &   0.1   \\
 & & & & & & & & & & \\
Total      &  21.5   &   2.8   &  13.7   &   1.6   &   1.8   &   0.6  
    &   0.5   &   0.4   &   0.2   &   0.3   \\
\hline
\end{tabular}
\end{center}
$^a$~All results are in units of antenna temperature.
Columns are the same as Table \ref{31a_syserr_table}.\\
\end{table}

\normalsize
\halfspace
\begin{table}
\caption{\label{53b_syserr_table}
Systematic Effects for Channel 53B$^a$}
\begin{center}
\begin{tabular}{l r r r r r r r r r r}
\hline
Effect & P-P & rms & $\Delta T_1$ & $\Delta T_2$ & $\Delta T_3$ 
       & $\Delta T_4$ & $\Delta T_5$ & $\Delta T_6$ & $\Delta T_7$ 
       & $\Delta T_8$ \\
       & ($\mu$K) & ($\mu$K) & ($\mu$K) & ($\mu$K) & ($\mu$K) & ($\mu$K) 
       & ($\mu$K) & ($\mu$K) & ($\mu$K) & ($\mu$K) \\
\hline
\multicolumn{11}{c}{Channel 53B Before Correction} \\
\hline
$\beta_X$  &   1.1   &   0.1   &   0.1   &   0.0   &   0.0   &   0.0  
    &   0.0   &   0.0   &   0.0   &   0.0   \\
$\beta_R$  &  31.5   &   5.3   &   2.8   &   3.2   &   1.2   &   3.6  
    &   0.9   &   0.8   &   0.2   &   0.5   \\
$\beta_T$  &  22.1   &   3.4   &  20.3   &   1.7   &   2.6   &   0.7  
    &   0.6   &   0.4   &   0.1   &   0.2   \\
Earth      &   6.6   &   0.8   &   0.6   &   0.7   &   0.2   &   0.2  
    &   0.1   &   0.1   &   0.0   &   0.1   \\
Moon       &   7.2   &   1.6   &   0.0   &   1.2   &   0.0   &   0.4  
    &   0.0   &   0.2   &   0.0   &   0.5   \\
Doppler    &  70.5   &  10.9   &  50.5   &   6.8   &   6.7   &   1.8  
    &   3.0   &   1.9   &   0.4   &   1.3   \\
Spin       &   2.1   &   0.3   &   1.0   &   0.1   &   0.2   &   0.1  
    &   0.1   &   0.1   &   0.0   &   0.0   \\
Other      &  41.9   &   4.3   &   2.7   &   1.3   &   0.4   &   0.6  
    &   0.3   &   0.5   &   0.4   &   0.5   \\
 & & & & & & & & & & \\
Total      &  91.2   &  13.4   &  54.6   &   7.9   &   7.3   &   4.1  
    &   3.2   &   2.2   &   0.6   &   1.5   \\
\hline
\multicolumn{11}{c}{Channel 53B After Correction} \\
\hline
$\beta_X$  &   0.2   &   0.0   &   0.0   &   0.0   &   0.0   &   0.0  
    &   0.0   &   0.0   &   0.0   &   0.0   \\
$\beta_R$  &  11.1   &   1.9   &   1.0   &   1.1   &   0.4   &   1.3  
    &   0.3   &   0.3   &   0.1   &   0.2   \\
$\beta_T$  &   6.4   &   1.0   &   5.8   &   0.5   &   0.8   &   0.2  
    &   0.2   &   0.1   &   0.0   &   0.1   \\
Earth      &   8.7   &   1.0   &   0.8   &   0.9   &   0.2   &   0.2  
    &   0.2   &   0.1   &   0.0   &   0.1   \\
Moon       &   0.9   &   0.2   &   0.0   &   0.2   &   0.0   &   0.1  
    &   0.0   &   0.0   &   0.0   &   0.1   \\
Doppler    &   1.2   &   0.2   &   0.7   &   0.1   &   0.1   &   0.0  
    &   0.1   &   0.0   &   0.0   &   0.0   \\
Spin       &   7.1   &   0.9   &   3.4   &   0.4   &   0.6   &   0.2  
    &   0.2   &   0.2   &   0.0   &   0.1   \\
Other      &  17.3   &   1.5   &   2.6   &   1.0   &   0.2   &   0.3  
    &   0.1   &   0.2   &   0.2   &   0.2   \\
 & & & & & & & & & & \\
Total      &  24.3   &   2.9   &   7.4   &   1.9   &   1.1   &   1.4  
    &   0.5   &   0.4   &   0.2   &   0.3   \\
\hline
\end{tabular}
\end{center}
$^a$~All results are in units of antenna temperature.
Columns are the same as Table \ref{31a_syserr_table}.\\
\end{table}

\normalsize
\halfspace
\begin{table}
\caption{\label{90a_syserr_table}
Systematic Effects for Channel 90A$^a$}
\begin{center}
\begin{tabular}{l r r r r r r r r r r}
\hline
Effect & P-P & rms & $\Delta T_1$ & $\Delta T_2$ & $\Delta T_3$ 
       & $\Delta T_4$ & $\Delta T_5$ & $\Delta T_6$ & $\Delta T_7$ 
       & $\Delta T_8$ \\
       & ($\mu$K) & ($\mu$K) & ($\mu$K) & ($\mu$K) & ($\mu$K) & ($\mu$K) 
       & ($\mu$K) & ($\mu$K) & ($\mu$K) & ($\mu$K) \\
\hline
\multicolumn{11}{c}{Channel 90A Before Correction} \\
\hline
$\beta_X$  &   4.9   &   0.4   &   1.6   &   0.1   &   0.2   &   0.1  
    &   0.1   &   0.0   &   0.0   &   0.0   \\
$\beta_R$  &  89.3   &  15.1   &   7.9   &   9.2   &   3.4   &  10.4  
    &   2.7   &   2.1   &   0.5   &   1.3   \\
$\beta_T$  &  37.8   &   5.9   &  35.0   &   2.9   &   4.5   &   1.1  
    &   1.0   &   0.6   &   0.2   &   0.4   \\
Earth      &   4.8   &   0.6   &   0.5   &   0.5   &   0.1   &   0.2  
    &   0.1   &   0.1   &   0.0   &   0.1   \\
Moon       &   8.4   &   1.8   &   0.1   &   1.4   &   0.0   &   0.5  
    &   0.0   &   0.3   &   0.0   &   0.6   \\
Doppler    &  62.3   &   9.6   &  44.3   &   5.9   &   5.9   &   1.6  
    &   2.7   &   1.6   &   0.4   &   1.1   \\
Spin       &   4.9   &   0.6   &   2.3   &   0.3   &   0.4   &   0.2  
    &   0.2   &   0.2   &   0.0   &   0.1   \\
Other      &  51.8   &   5.3   &   1.7   &   0.9   &   0.3   &   0.5  
    &   0.3   &   0.5   &   0.3   &   0.4   \\
 & & & & & & & & & & \\
Total      & 126.9   &  19.7   &  57.1   &  11.4   &   8.2   &  10.6  
    &   3.9   &   2.8   &   0.8   &   2.0   \\
\hline
\multicolumn{11}{c}{Channel 90A After Correction} \\
\hline
$\beta_X$  &   1.0   &   0.1   &   0.3   &   0.0   &   0.0   &   0.0  
    &   0.0   &   0.0   &   0.0   &   0.0   \\
$\beta_R$  &  12.3   &   2.1   &   1.1   &   1.3   &   0.5   &   1.4  
    &   0.4   &   0.3   &   0.1   &   0.2   \\
$\beta_T$  &  10.3   &   1.6   &   9.5   &   0.8   &   1.2   &   0.3  
    &   0.3   &   0.2   &   0.1   &   0.1   \\
Earth      &   9.4   &   1.2   &   1.0   &   1.0   &   0.2   &   0.3  
    &   0.2   &   0.2   &   0.1   &   0.1   \\
Moon       &   1.4   &   0.3   &   0.0   &   0.2   &   0.0   &   0.1  
    &   0.0   &   0.1   &   0.0   &   0.1   \\
Doppler    &   2.6   &   0.4   &   1.8   &   0.3   &   0.2   &   0.1  
    &   0.1   &   0.1   &   0.0   &   0.1   \\
Spin       &  12.1   &   1.6   &   5.8   &   0.7   &   1.1   &   0.4  
    &   0.4   &   0.4   &   0.0   &   0.2   \\
Other      &  15.4   &   1.0   &   1.6   &   0.7   &   0.1   &   0.2  
    &   0.1   &   0.2   &   0.1   &   0.2   \\
 & & & & & & & & & & \\
Total      &  27.2   &   3.5   &  11.5   &   2.0   &   1.7   &   1.6  
    &   0.7   &   0.6   &   0.2   &   0.4   \\
\hline
\end{tabular}
\end{center}
$^a$~All results are in units of antenna temperature.
Columns are the same as Table \ref{31a_syserr_table}.\\
\end{table}

\normalsize
\halfspace
\begin{table}
\caption{\label{90b_syserr_table}
Systematic Effects for Channel 90B$^a$}
\begin{center}
\begin{tabular}{l r r r r r r r r r r}
\hline
Effect & P-P & rms & $\Delta T_1$ & $\Delta T_2$ & $\Delta T_3$ 
       & $\Delta T_4$ & $\Delta T_5$ & $\Delta T_6$ & $\Delta T_7$ 
       & $\Delta T_8$ \\
       & ($\mu$K) & ($\mu$K) & ($\mu$K) & ($\mu$K) & ($\mu$K) & ($\mu$K) 
       & ($\mu$K) & ($\mu$K) & ($\mu$K) & ($\mu$K) \\
\hline
\multicolumn{11}{c}{Channel 90B Before Correction} \\
\hline
$\beta_X$  &   0.1   &   0.0   &   0.0   &   0.0   &   0.0   &   0.0  
    &   0.0   &   0.0   &   0.0   &   0.0   \\
$\beta_R$  &  10.7   &   1.8   &   0.9   &   1.1   &   0.4   &   1.2  
    &   0.3   &   0.3   &   0.1   &   0.2   \\
$\beta_T$  &  15.4   &   2.4   &  14.2   &   1.2   &   1.8   &   0.5  
    &   0.4   &   0.3   &   0.1   &   0.2   \\
Earth      &   5.3   &   0.6   &   0.5   &   0.5   &   0.1   &   0.2  
    &   0.1   &   0.1   &   0.0   &   0.1   \\
Moon       &   9.4   &   2.0   &   0.1   &   1.5   &   0.0   &   0.6  
    &   0.0   &   0.3   &   0.0   &   0.6   \\
Doppler    &  62.7   &   9.6   &  44.3   &   5.9   &   5.9   &   1.6  
    &   2.7   &   1.6   &   0.5   &   1.2   \\
Spin       &   0.1   &   0.0   &   0.0   &   0.0   &   0.0   &   0.0  
    &   0.0   &   0.0   &   0.0   &   0.0   \\
Other      &  43.1   &   4.3   &   1.4   &   0.8   &   0.3   &   0.5  
    &   0.3   &   0.4   &   0.3   &   0.4   \\
 & & & & & & & & & & \\
Total      &  79.1   &  11.1   &  46.5   &   6.4   &   6.2   &   2.2  
    &   2.7   &   1.8   &   0.6   &   1.4   \\
\hline
\multicolumn{11}{c}{Channel 90B After Correction} \\
\hline
$\beta_X$  &   0.5   &   0.0   &   0.1   &   0.0   &   0.0   &   0.0  
    &   0.0   &   0.0   &   0.0   &   0.0   \\
$\beta_R$  &   2.9   &   0.5   &   0.3   &   0.3   &   0.1   &   0.3  
    &   0.1   &   0.1   &   0.0   &   0.0   \\
$\beta_T$  &   6.7   &   1.0   &   6.2   &   0.5   &   0.8   &   0.2  
    &   0.2   &   0.1   &   0.0   &   0.1   \\
Earth      &   7.8   &   0.9   &   0.7   &   0.8   &   0.2   &   0.2  
    &   0.1   &   0.1   &   0.0   &   0.1   \\
Moon       &   1.5   &   0.3   &   0.0   &   0.3   &   0.0   &   0.1  
    &   0.0   &   0.1   &   0.0   &   0.1   \\
Doppler    &   1.8   &   0.3   &   1.2   &   0.2   &   0.2   &   0.1  
    &   0.1   &   0.1   &   0.0   &   0.0   \\
Spin       &   5.4   &   0.7   &   2.6   &   0.3   &   0.5   &   0.2  
    &   0.2   &   0.2   &   0.0   &   0.1   \\
Other      &  11.2   &   0.8   &   1.3   &   0.5   &   0.2   &   0.2  
    &   0.1   &   0.1   &   0.1   &   0.1   \\
 & & & & & & & & & & \\
Total      &  16.6   &   1.9   &   7.0   &   1.2   &   1.0   &   0.5  
    &   0.3   &   0.3   &   0.1   &   0.2   \\
\hline
\end{tabular}
\end{center}
$^a$~All results are in units of antenna temperature.
Columns are the same as Table \ref{31a_syserr_table}.\\
\end{table}

\clearpage

\end{document}